\documentclass[reprint,aps,prb,noeprint]{revtex4-2}
\usepackage{amsmath,amssymb,mathtools,tensor,braket}
\usepackage{bm,color}
\usepackage[pdfusetitle]{hyperref}
\hypersetup{pdfauthor={Guihyun Han, Minkyu Park, Sung Hyon Rhim}}
\usepackage[all]{hypcap}

\begin{document}
%======================
% For line numbers
%=============
%\linenumbers
%\modulolinenumbers[5] 
%=================================
%===============================================
%       TITLE
%=================================================
\title{Strain tunable anomalous Hall and Nernst conductivities in compensated ferrimagnetic Mn$_3$Al} 
%===============================================
%                                        AUTHORs
%================================================
\author{Guihyun \surname{Han}$^1$} 
\author{Minkyu \surname{Park}$^{1,2}$ }
\email{minkyupark@kriss.re.kr}
\author{S. H. \surname{Rhim}$^1$}
\email{sonny@ulsan.ac.kr}
%===============================================
%                                    AFFILIATION
%===============================================
\affiliation{
$^1$ Department of Physics, University of Ulsan, 93, Daehak-ro, Nam-gu, Ulsan, Republic of Korea
\\ $^2$ Quantum Technology Institute, Korea Research Institute of Standards and Science, 267, Gajeong-ro, Yuseong-gu, Daejeon, Republic of Korea
}
%==============================================
%                                        ABSTRACT
%==============================================
\begin{abstract}
The tunability of anomalous Hall and Nernst conductivities is investigated in the compensated ferrimagnet Mn$_3$Al 
under isotropic strain ($\eta$) and chemical potential variation using first-principles calculations. 
At a chemical potential of $\mu = -0.3$ eV, three distinct topological features---Weyl points, nodal lines, and gapped nodal lines---are simultaneously realized along high-symmetry directions of the Brillouin zone in the framework of magnetic space group.
The anomalous Hall conductivity (AHC) is found to be predominantly governed by the Berry curvature in the $k_y k_z$ plane 
and can be enhanced significantly under tensile strain, reaching $-1200$ $(\Omega~\mathrm{cm})^{-1}$. 
On the other hand, the anomalous Nernst conductivity (ANC) shows a sign change near the Fermi level and 
whose magnitude increases at $\mu = -0.3$ eV with quasi-quadratic strain dependence. 
Regardless of strain, the underlying bands and Fermi surface structures remain robust, 
while the distribution and magnitude of Berry curvature evolve substantially. 
These results underscore the potential of Mn$_3$Al, a compensated ferrimagnet, 
as a platform for Berry curvature engineering via strain and doping.
\end{abstract}
%==============================================
\maketitle
%==============================================
%.                            SEC. 0 Introduction
%==============================================

\section{Introduction}

In modern spintronics, the role of the anomalous Hall effect (AHE) and anomalous Nernst effect (ANE) are increasing,
by providing an efficient mechanism to generate transverse charge and spin current 
without relying on external magnetic fields\cite{Nakatsuji:2015aa,Kimata:2019aa,Uchida:nat2008}. 
Both AHE and ANE arise from the interplay of topology of electronic structure, spin-orbit coupling, 
and broken time-reversal symmetry \cite{Nagaosa_2010,PhysRevLett.112.017205,K_bler_2014,PhysRevLett.88.207208,PhysRevX.12.011028}.
The intrinsic origins of AHE and ANE are now widely understood in terms of the Berry curvature of electronic bands, with topological features such as Weyl points and nodal lines playing crucial roles\cite{PhysRevLett.97.026603,RevModPhys.82.1959,Nagaosa_2010}.

In recent years, compensated ferrimagnets and some antiferromagnets, with vanishing net magnetization,
have emerged as promising hosts of large AHE and ANE,  
combining broken time-reversal symmetry with vanishing net magnetization\cite{Yao_2004,K_bler_2014,PhysRevX.12.011028,ZFang:Sci-2003,Smejkal:2022aa,RevModPhys.90.015005,Jungwirth:2016aa}. 
The vanishing net magnetization is advantageous for overcoming stray fields while still supporting a robust Berry curvature. 
In particular, Heusler compounds are versatile owing to tunable crystal structures, rich magnetic phase diagrams, 
and compatibility with spintronic technologies\cite{GRAF20111,PhysRevB.85.012405,Wollmann-ann.rev.mat:2017}.
Such flexibility makes Heusler compounds excellent platforms for studying and engineering topological transport properties.

A representative example is Mn$_3$Al, a compensated collinear ferrimagnet in a cubic Heusler, 
with half-metallicity and Curie temperature of 605 K\cite{Jamer_2017}.
The magnetic symmetry of Mn$_3$Al allows a nonzero anomalous Hall conductivity despite zero net magnetization\cite{PhysRevResearch.4.013215,PhysRevB.97.060406}. 
Previous studies have identified its intrinsic AHE, associated with symmetry-protected Berry curvature hot spots\cite{PhysRevResearch.4.013215}. However, how this anomalous response evolves under external tuning---such as isotropic strain or chemical potential variation---has not been fully clarified. 
Moreover, the possible coexistence of multiple topological features under such conditions has remained unexplored.

In this work, we employ first-principles calculations to reveal that Mn$_3$Al simultaneously hosts Weyl points, nodal lines, and gapped nodal lines around a chemical potential of $\mu=-0.3$ eV, located along high-symmetry directions of its magnetic Brillouin zone. 
These topological features give rise to strongly strain-dependent variations in both anomalous Hall and Nernst conductivities. 
Our findings suggest that compensated Heusler ferrimagnets provide 
a promising route for Berry-curvature engineering of spintronic devices\cite{PhysRevB.95.075128,Smejkal:2020fg,PhysRevB.99.165117,Li:2020uj,PhysRevB.103.085116,PhysRevResearch.2.043366,PhysRevMaterials.4.051401,PhysRevB.97.214402,https://doi.org/10.1002/advs.202201749,Son:2016aa}.
Such engineering has already been demonstrated under hydrostatic pressure \cite{PhysRevMaterials.4.051401,PhysRevResearch.2.043366,PhysRevB.97.214402} and via chemical potential variation \cite{https://doi.org/10.1002/advs.202201749,Son:2016aa}.
%==============================================
%.                SEC. 1 Structure and calculations
%==============================================
\section{Structure and magnetic space group}\label{sec:MSG}
%==============================================
%.                FIG 1 Structure & BZ
%==============================================
\begin{figure}[b]
\includegraphics[width=\columnwidth]{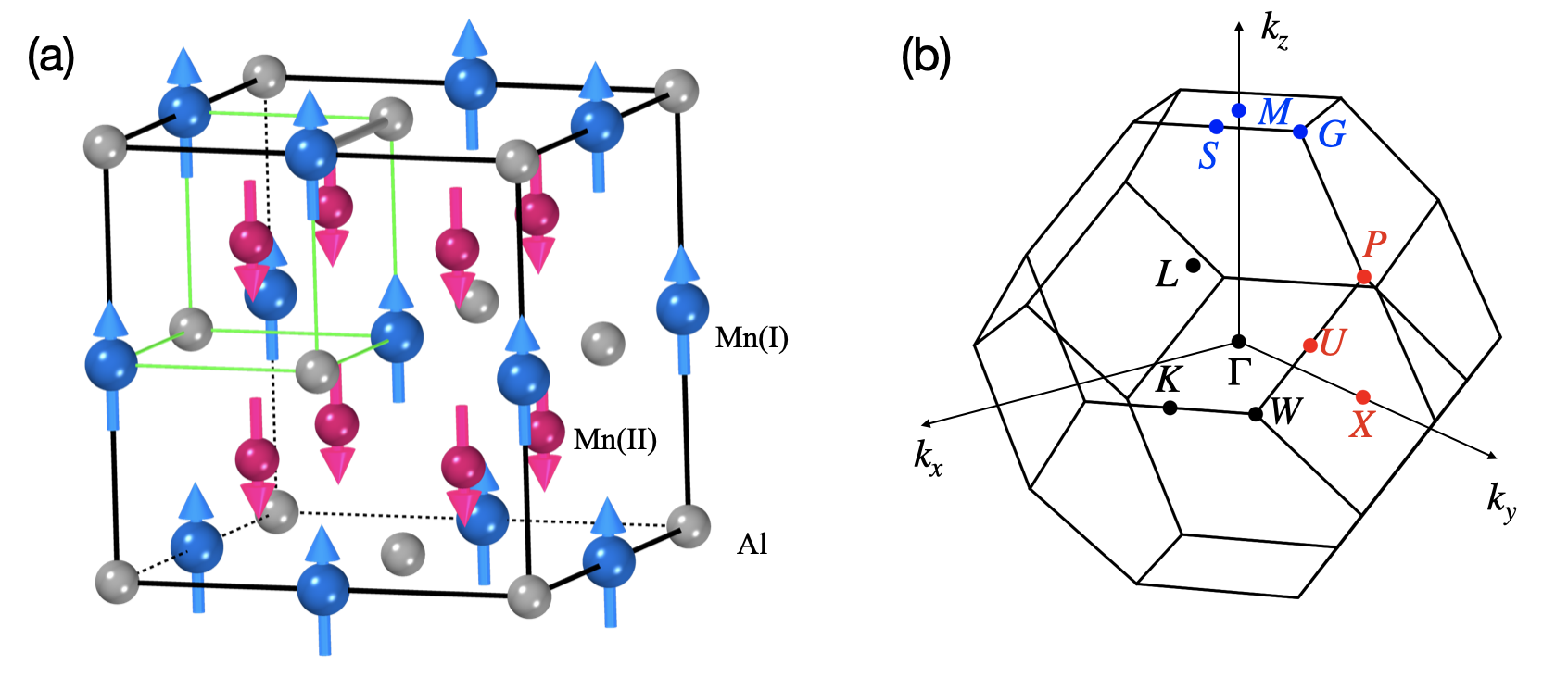}
\caption{(a) Crystal structure of Mn$_3$Al with magnetization along [$001$]. 
Mn(I), Mn(II), and Al are in blue, red, and gray spheres, respectively,
with magnetic moments shown in arrows.
Green line denotes cube formed by Mn(I) and Al, where Mn(II) is at the center of the cube.
(b) Brillouin zone of Mn$_3$Al of magnetic space group $I4/mm'm'$ (No. 139.537).
High symmetry points, symmetrically equivalent in space group but  distinct in magnetic space group, are distinguished by colors.  
}
\label{fig:1}
\end{figure}

Crystal structure of a regular Heusler compound, Mn$_3$Al, in  $D0_3$ structure is depicted in Fig.~\ref{fig:1} (a),
whose space group is $Fm\bar{3}m$ (No.~225). 
It can be viewed as a superposition of two $fcc$ of Al and Mn(I) with Mn(II) locating at the center of cubic box formed by Mn(I) and Al (denoted by green lines).
Site symmetry of constituent atoms are $m\bar{3}m(4a)$ for Al, $m\bar{3}m(4b)$ for Mn(I), and $\bar{4}3m(8c)$  for Mn(II), respectively. 
Different site symmetry results in distinct magnetic moments of Mn(I) and Mn(II).
While Al has negligible moment, moment of Mn(I) is $2.6\mu_B$ and that of Mn(II) is $-1.3\mu_B$, respectively.

Brillouin zone of $\mathrm{Mn_3Al}$ is shown in Fig.~\ref{fig:1} (b) with labels for high symmetry points.
For $Fm\bar{3}m$, high-symmetry points shown in blue $(M, S, G)$ and red $(X, U, P)$ are equivalent. 
They are inequivalent in the presence of magnetization along $z$ axis, whose magnetic space group is $I4/mm'm'$ (No.~139.537). 
Under 4-fold rotation along $z$ axis, $4_{001}$, the crystal structure is invariant. 
On the other hand, the fourfold rotation $4_{100}$ around the $x$ axis leaves the chemical structure invariant but alters the magnetic structure. 
Hence, $4_{001}$ is the symmetry operation of magnetic space group, $I4/mm'm'$ (No. 139.537), but $4_{100}$ is not. 
In this context, $\Gamma-M$ has 4-fold rotational symmetry while $\Gamma-X$ has no such in $I4/mm'm'$ (No. 139.537).

\section{\label{sec:calc}Methods of Calculations}
First-principles calculations are performed using Vienna \emph{ab initio} simulation package (VASP) with spin-orbit coupling included\cite{vasp}. 
For the exchange-correlation energy, generalized gradient approximation (GGA) is used with Perdew-Burke-Ernzerhof  (PBE) parameterization \cite{PhysRevLett.77.3865}.
Summation in Brillouin zone is done with $15\times15\times15$ $k$ grid.
Cutoff energy for wave function expansion is $450 \; \mathrm{eV}$.
Optimized lattice constant is $5.78 \; \AA$ in good agreement with experiment \cite{Jamer_2017}.

After self-consistent calculations, maximally-localized Wannier functions (MLWFs) are obtained using Wannier90 \cite{Wang_2006,PhysRevB.65.035109} with $9\times9\times9$ $k$ grid.
The initial projections were chosen using $d$ orbitals for Mn and $s$ and $p$ for Al atoms.
38 maximally-localized Wannier functions (MLWFs) are constructed from 72 Bloch bands.
Berry curvature is computed using WannierBerri \cite{Tsirkin:2021aa}via Wannier interpolation 
with $300\times300\times300$ $k$ grid.

Anomalous Hall conductivity (AHC) is obtained 
by summing up Berry curvature, $\Omega_{n,\alpha\beta}(\bf{k})$, over occupied states in Brillouin zone\cite{Yao_2004}, 
\begin{equation}
\sigma_{\alpha\beta}=\frac{e^2}{\hbar}\int \frac{d^3k}{(2\pi)^3}\sum_n f_{n\mathbf{k}}{\Omega_{n,\alpha\beta}(\bf{k})},
\label{eq:ahc}
\end{equation}
where $e$ is the elementary charge; $f_{n\textbf{k}}$ is the Fermi-Dirac distribution of $n$-th  band of $k$.
The Berry curvature is obtained from Kubo formula,
\begin{equation}
\Omega_{\alpha \beta}(\mathbf{k})=-\sum_{n\neq n'}\frac{2\mathrm{Im}\left[ \langle n\mathbf{k}  | \mathrm{v}_{\alpha}| n' \mathbf{k}\rangle \langle n'\mathbf{k}  | \mathrm{v}_{\beta}| n\mathbf{k}\rangle \right]}{(\varepsilon_{n\mathbf{k}}-\varepsilon_{n'\mathbf{k}})^2},
\label{eq:bc}
\end{equation}
where $\mathrm{v}_{\alpha}$ is the velocity operator of $\alpha$-th component; 
$\varepsilon_{n{\bf k}}$ is energy of $n$-th band of $k$.

Anomalous Nernst conductivity (ANC) is expressed in terms of Berry curvature\cite{PhysRevLett.97.026603},
\begin{equation}
\alpha_{\alpha\beta}=\frac{e}{\hbar}\int  \frac{d^3k}{(2\pi)^3}\sum_n S_n(\mathbf{k}){\Omega_{n,\alpha\beta}(\mathbf{k})}.
\label{eq:anc}
\end{equation}
Here $S_n({\bf k})$ is the entropy of fermion,
\begin{equation}
S_n({\bf k})=-k_B[f_{n{\bf k}}\mathrm{log}f_{n{\bf k}}+(1-f_{n{\bf k}})\mathrm{log}(1-f_{n{\bf k}})],
\label{eq:S}
\end{equation}
where $k_B$ is the Boltzmann constant.
In the low temperature limit, ANC is commonly expressed by Mott relation,
$\alpha_{\alpha\beta}=({\pi^2}/{3})({k_B^2T}/{e})\lim_{T\rightarrow 0} {\partial \sigma_{\alpha\beta}}/{\partial \mu}$.
This limit represents the Berry curvature contribution on the Fermi surface\cite{PhysRevLett.97.026603,PhysRevResearch.4.013215}.

\section{\label{sec:res}Results and Discussions}

\subsection{\label{sec:symm}Band crossing and Nodal line}

%==============================================
%. Fig.2 band structure and Brillouin zone
%===============================================
\begin{figure*}[htbp]
\centering
\includegraphics[width=0.75\textwidth]{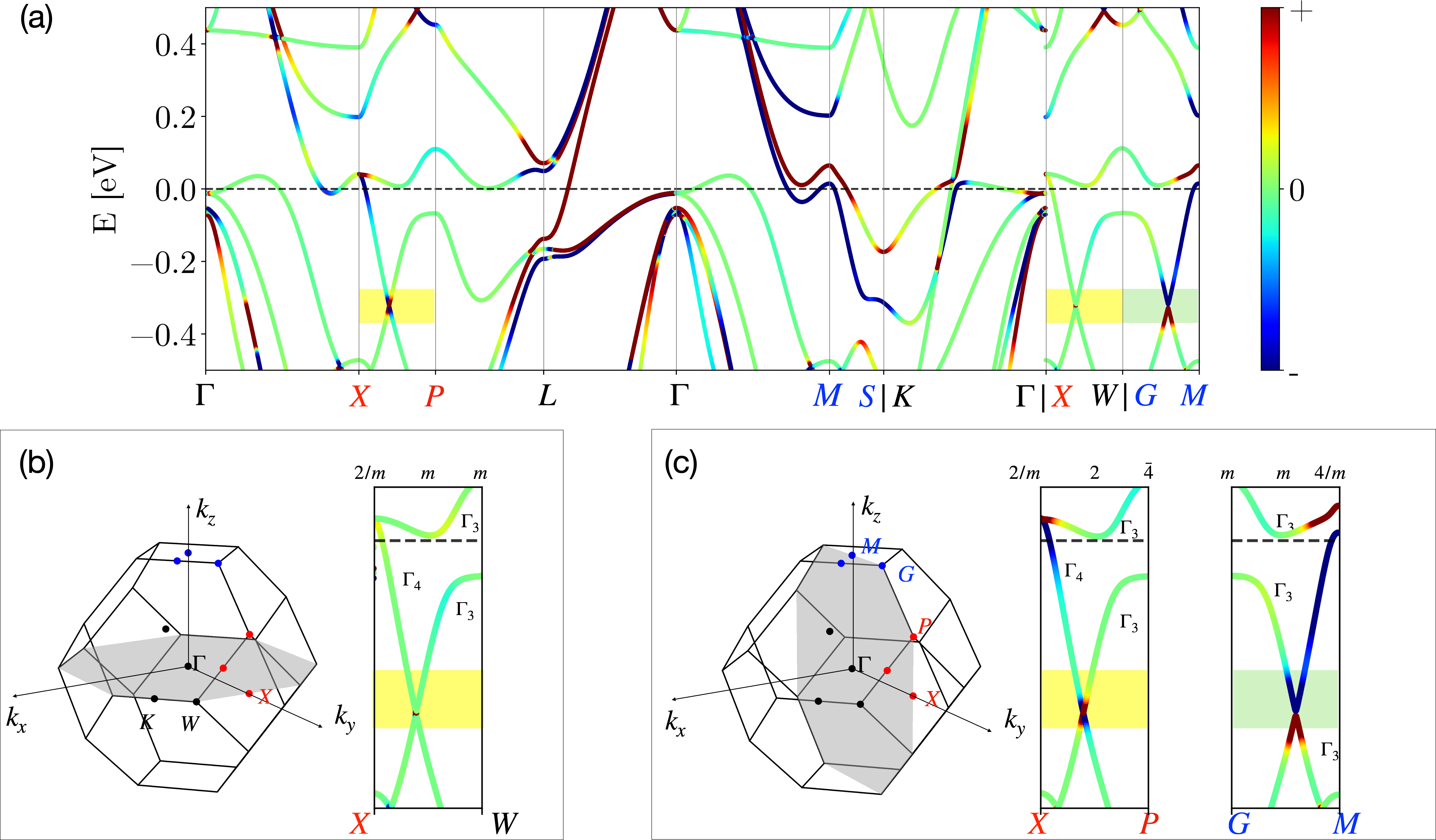}
\caption{
(a) $k$ projected Berry curvature of Mn$_3$Al without strain. 
Symmetrically distinct $k$ points are distinguished by colors in black, blue, and red.
Shade boxes denote: crossing  (in yellow) and avoided crossing  (in blue).
Crossing are further classified into crossing and Weyl points (see text).
(b) (001) plane in Brillouin zone containing $X-W$, where crossing is denoted in yellow box.
(c) (100) plane  in Brillouin zone containing  $X-P$ and $G-M$, which are Weyl point and avoided crossing, respectively.
In insets of (b) and (c), group of $k$ and representation of bands are shown, where color of bands are $k$ resolved Berry curvature.
}
\label{fig:2}
\end{figure*}
%===========================================
Band structure of Mn$_3$Al without strain is shown in Fig.~\ref{fig:2}.  
Red (blue) of bands represent the amplitude of $k$ resolved Berry curvature with positive (negative) sign.
%In Fig.~\ref{fig:2} (a), color-labeled high symmetry points are consistent with those in Fig.~\ref{fig:1}(b). 
In Fig.~\ref{fig:2} (a), color-labeled high symmetry points are classified as crossing, Weyl points, and avoided crossing, respectively.
Mn$_3$Al is half-metallic with band gap 0.4 eV in the minority spin channel. 
This half-metallicity is robust with strain \footnote{See Figs. S1 and S2 of Supplemental Material at http://link.aps.org/supplemental/xxxxx . Spin resolved band structure with and without strains are shown. At $\Gamma$, the half-metallicity is robust while the size of band gap changes.}.
Around $\Gamma$, large $\Omega$ emerges with alternating signs. 
More dominant $\Omega$ is evident near $L$ : while bands around $E_F-0.2$ eV have both signs,
rather dispersive band crossing around $E_F$ has large contribution. 
$\Omega$ around $M$ has also pronounced contribution near $E_F$ with both signs.

Fig.~\ref{fig:2} (b-c) show enlarged views of colored boxes of Fig.~\ref{fig:2} (a).
Brillouin zone with ($001$) and ($100$) plane are shown respectively, which contain
$k$ paths of our interest. Band analysis along $X-W$ is shown in Fig.~\ref{fig:2} (b); 
that along $X-P$ and $G-M$ is in Fig.~\ref{fig:2} (c).
Little groups with corresponding representations are on top 
along color-coded Berry curvature.

Along $X-W$, band crossing of $XW_3$ and $XW_4$ states are evident,
where ($001$) is the mirror plane containing $X-W$.
Along $X-P$, the band crossing of $XP_3$ and $XP_4$ is Weyl point,
where ($100$) is not a mirror plane when the magnetization is along ($001$). 
On the other hand, the path along $G-M$, 
the ($100$) plane not being a mirror plane results in avoided crossing of $GM_3$ and $GM_3$,
where positive and negative Berry curvature is apparent above and below the gap.
From this symmetry analysis, Mn$_3$Al contains band crossing, Weyl point, and avoided crossing when $\mu=E_F - 0.3$ eV
\footnote{See Fig. S8 of Supplemental Material at http://link.aps.org/supplemental/xxxxx. For this chemical potential shift, negative charge of 1.52 is needed.}.

\subsection{\label{sec:AHC-ANC}Anomalous Hall and Nernst Conductivity}
\begin{figure}[b]
\centering
\includegraphics[width=\columnwidth]{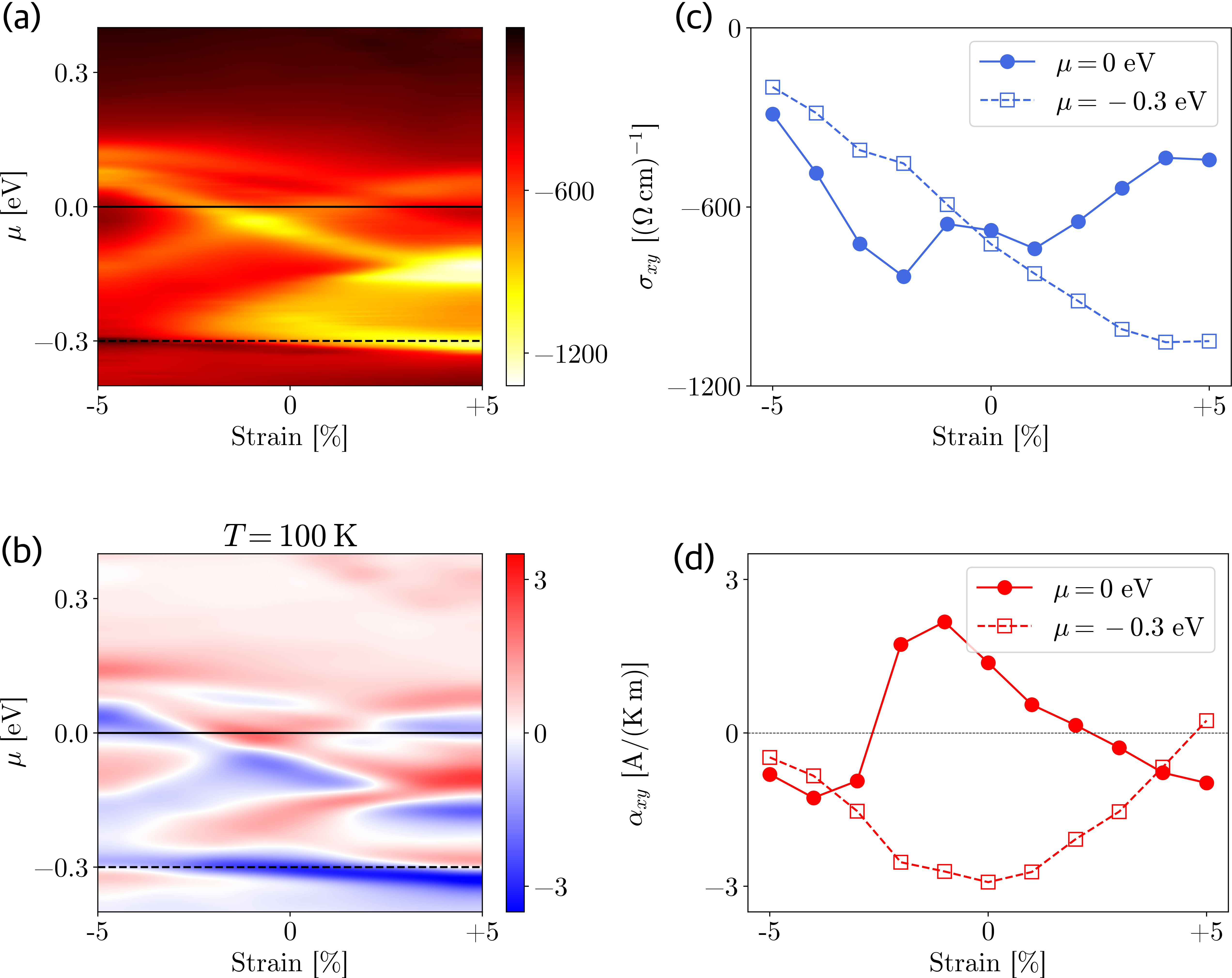}
\caption{ 
(a) Anomalous Hall conductivity ($\sigma_{xy}$) and (b) anomalous Nernst conductivity ($\alpha_{xy}$) 
as function of strain and chemical potential in color contour.
Solid and dashed lines correspond to $\mu=E_F$ and $\mu=E_F - 0.3$ eV, respectively. 
(c) $\sigma_{xy}$ and (d) $\alpha_{xy}$ with respect to strain 
for $\mu=E_F$ (solid line with filled symbol) and $\mu=E_F - 0.3$ eV (dashed line with open symbols), respectively. 
}
\label{fig:3}
\end{figure}

Fig.~\ref{fig:3} presents the anomalous Hall ($\sigma_{xy}$) and the Nernst conductivity ($\alpha_{xy}$). %with respect to strains and the chemical potential ($\mu$).
In Fig.~\ref{fig:3} (a-b), $\sigma_{xy}$ and $\alpha_{xy}$ are presented 
as function of strain ($\eta$) and the chemical potential ($\mu$) in color contour, where bright color correspond large $|\Omega|$.
Solid (dashed) lines denote $\mu=E_F$ ($\mu=E_F-0.3$ eV), where $T=100~K$ is used.
For $\sigma_{xy}$, bright yellow region appears for $\eta=-5\%$ just above $\mu=E_F$,
which rather monotonically shifts to positive strain with $\mu<E_F$. 
Bright area is most pronounced for positive strains when $E_F-0.3 < \mu < E_F-0.1$ eV.

%===========================================
% AHE and ANE
%===========================================
%===========================================
%-- FIG3 - STATEMENT
%---- 재활용할 부분-----
%---------
For $\alpha_{xy}$, positive (negative) values are denoted by red (blue) colors.
As seen, when $\mu=E_F$, both positive and negative $\alpha_{xy}$ appear with strain.
Without strain $\sigma_{xy} = 670~(\Omega~\mathrm{cm})^{-1}$,
which becomes largest with 830 $(\Omega~\mathrm{cm})^{-1}$ for $\eta=-2\%$;
%When strain just less than 0\%, $\alpha_{xy}$ around $\mu=E_F$ is large with positive sign,
while $\alpha_{xy}$ around $\mu=E_F-0.3$ eV is large around 2.92~$(\mathrm{A/K~m})$ with negative sign.

To reveal clear strain dependence, 
$\sigma_{xy}$ and $\alpha_{xy}$ are plotted in Fig.~\ref{fig:3} (c-d) as function of strain,
for two chemical potentials, $\mu=E_F$ and $\mu=E_F - 0.3$ eV, denoted by solid and open symbols, respectively.
First, AHC is plotted in Fig.~\ref{fig:3} (c). 
When $\mu=E_F$, $\sigma_{xy}$, is in the range of $-1200$ $\sim$ $-300$ $(\Omega \mathrm{cm})^{-1}$.
%No apparently distinct feature is found with respect to strain.
For $\eta=+2\%$, $\sigma_{xy}$ is largest with -800 $(\Omega \mathrm{cm})^{-1}$.
%while for tensile strain $|\sigma_{xy}|$ tends to increase. 
When $\mu=E_F-0.3$ eV, on the other hand, $\sigma_{xy}$ exhibits clear tendency of monotonic decrease, 
or increase of magnitude, which reaches as high as $-1100$ $(\Omega \mathrm{cm})^{-1}$ for $\eta=+5\%$.
Second, ANC is plotted in Fig.~\ref{fig:3} (d). 
When $\mu=E_F$, $\alpha_{xy}$ is negative for compressive strain turning positive for $\eta=-2\%$.
It becomes negative for  $\eta>+3\%$.
$\alpha_{xy}$ reaches 2.17 and 1.73~$(\mathrm{A/K~m})$ for $\eta=0\%$ and $\eta=-1\%$.
The sign change of $\alpha$, as expressed in Eq.\ref{eq:anc},  is a manifest of sign of over all Berry curvature with strain.
%\sonny{Need $\frac{\partial \sigma}{\partial \mu}$ in method for discussion?}
When $\mu=E_F-0.3$ eV, $\alpha_{xy}$ exhibits quasi-quadratic feature with local minimum $-2.92$~$(\mathrm{A/K~m})$ near zero strain
\footnote{Similar feature is found for opposite chemical potential shift, $\mu=E_F+0.12$ eV, where the magnitude of AHC monotonically decreases, or $\sigma_{xy}$ increases from -680 to -292 S/cm with strain. This is not by topological singularity but by band splitting associated with representation near $L$. See Sec. V of Supplemental Material at
  http://link.aps.org/supplemental/xxxxx.}.

%================================================
% Section. volume evolution of BC on plane
%================================================
%\section{Fermiology}
%\subsection{Fermiology: $\mu=E_F$ and $\mu=E_F-0.3$ eV}
\subsection{Fermiology: $\mu=E_F$ and $\mu=E_F-0.3$ eV}
\label{sec:FS}
\begin{figure}[t]
\centering
\includegraphics[width=\columnwidth]{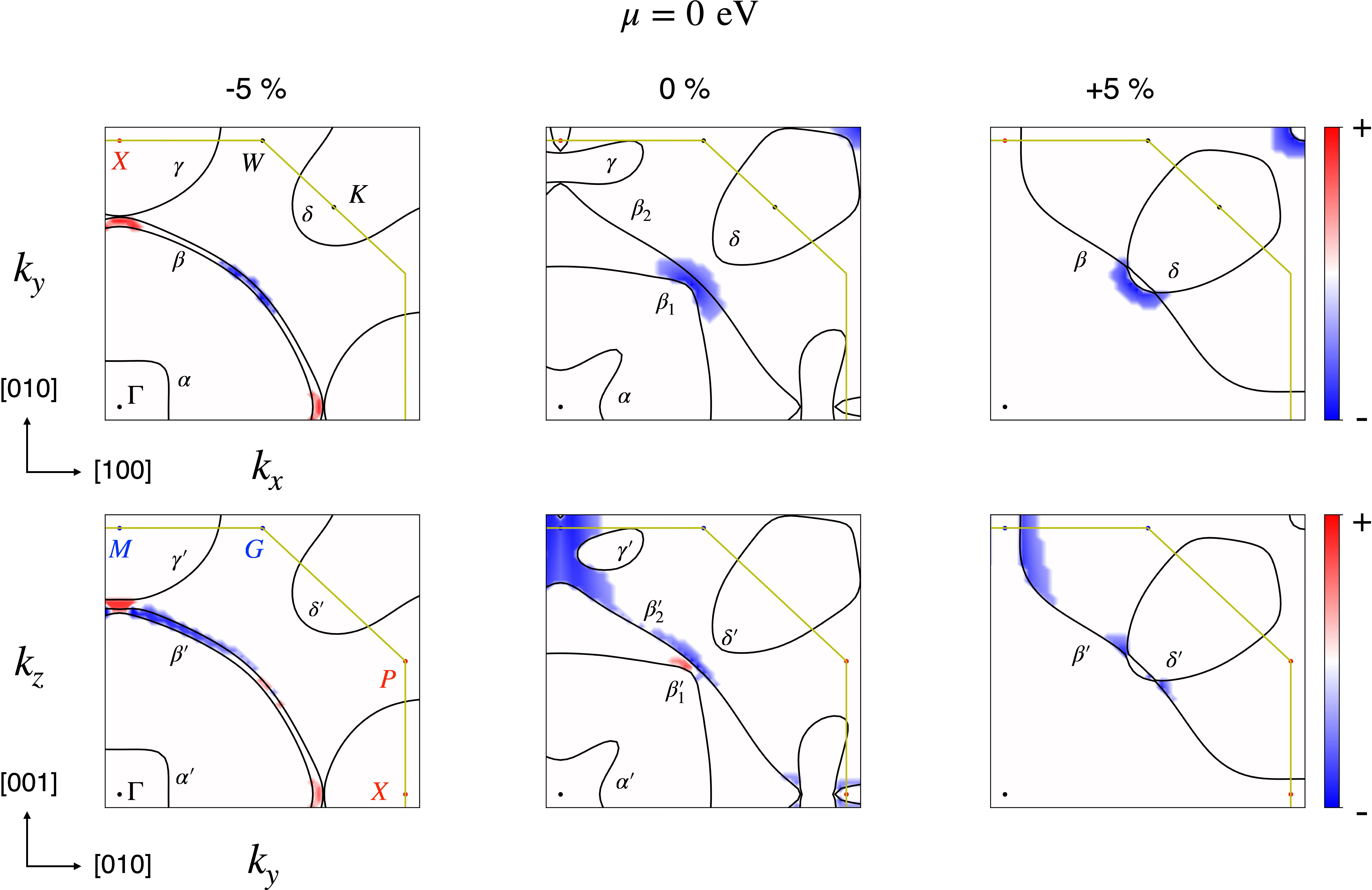}
\caption{ 
Contour of Berry curvature when $\mu=E_F$ for $\eta=-5$, $0$, and $+5\%$ from left to right column.
Upper panel: $\Omega_{xy}({\bf k})$ on $k_xk_y$ plane ($001$) and 
lower panel: contour on $k_yk_z$ plane ($100$).
Red (blue) color denotes positive (negative) $\Omega_{xy}({\bf k})$. Yellow line is for Brillouin zone boundary.
Fermi surface is shown in black lines, where individual Fermi sheets are labeled by $\alpha$, $\beta$, $\gamma$, and $\delta$.
High symmetry point labeled in black, red, and blue are consistent with Fig.~\ref{fig:2}.
}
\label{fig:4}
\end{figure}

\begin{figure}[t]
\centering
\includegraphics[width=\columnwidth]{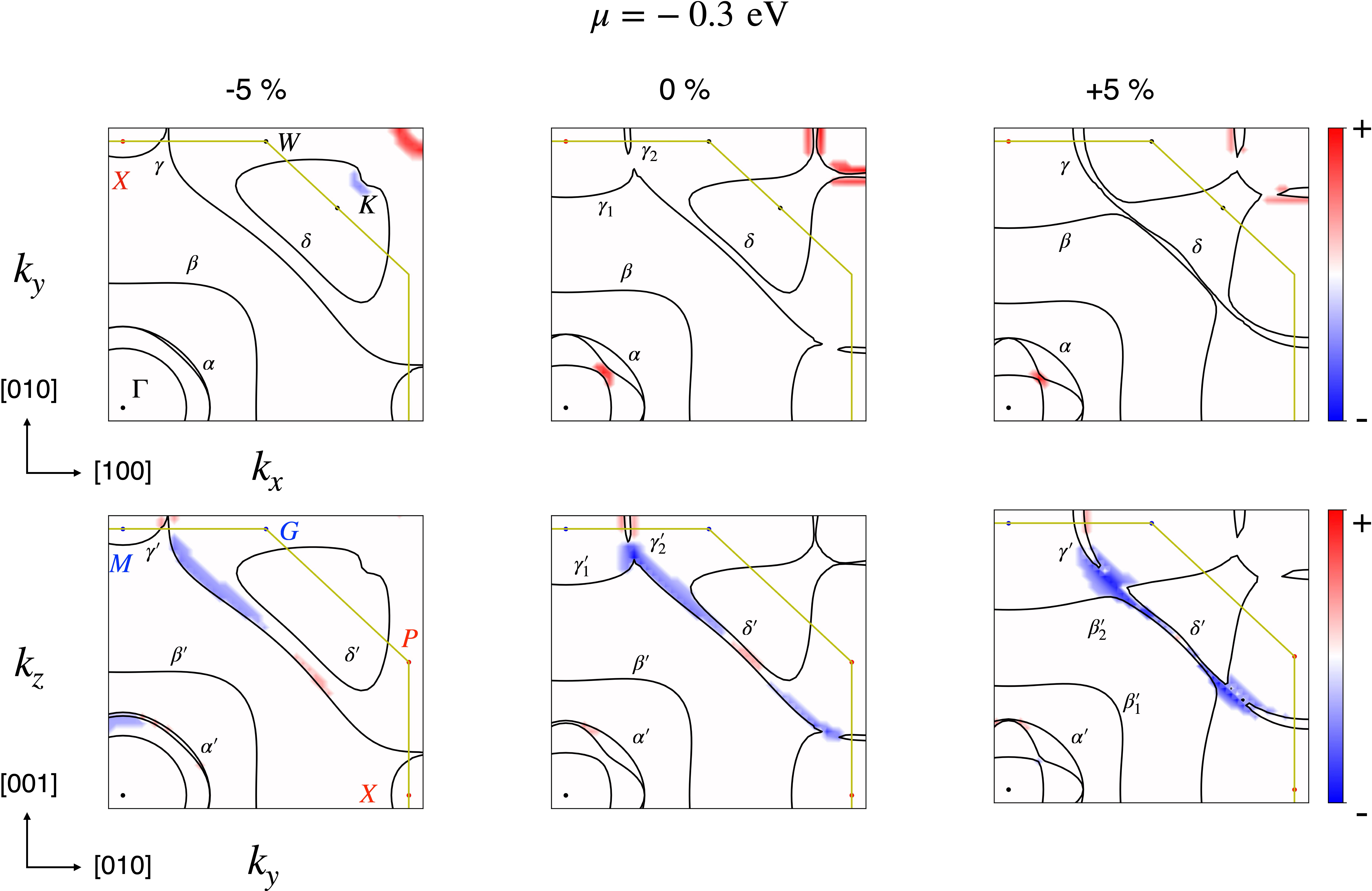}
\caption{ 
Contour of Berry curvature when $\mu=E_F-0.3$ eV for $\eta=-5$, $0$, and $+5\%$.
Upper panel: $\Omega_{xy}({\bf k}) $ near the $k_xk_y$ plane ($001$) and 
lower panel: contour on $k_yk_z$ plane ($100$).
%From left to right: strains of -5\%, 0, and +5\%.
Red (blue) color denotes positive (negative) $\Omega_{xy}({\bf k})$. Yellow line is for Brillouin zone boundary.
Fermi surface is shown in black lines, where individual Fermi sheets are labeled by $\alpha'$, $\beta'$, $\gamma'$, and $\delta'$.
High symmetry point labelled in black, red, and blue are consistent with Fig.~\ref{fig:2}.
}
\label{fig:5}
\end{figure}

As discussed earlier, AHC and ANC exhibit differently for $\mu=E_F$ and $\mu=E_F-0.3$ eV.
When $\mu=E_F$,  the Berry curvature is maximized without strain both on $k_xk_y$ and $k_yk_z$ planes.
With compressive strain, the magnitude of AHC is reduced with positive Berry curvature. 
With tensile strain, AHC is reduced with the smaller magnitude of Berry curvature in narrow region.
When $\mu=E_F-0.3$ eV, Berry curvature is dominantly from $k_yk_z$ plane.
The magnitude of Berry curvature increases with strain.
Here, the evolution of Fermi surface (FS) with respect to strain is discussed with concomitant distribution of Berry curvature for each $\mu$.
Fig.~\ref{fig:4} and Fig.~\ref{fig:5} show Berry curvature distribution with FS contour on $k_xk_y$ and $k_yk_z$ planes, 
whose bands structures are presented in Fig.~\ref{fig:2} (b) and (c). 
With SOC, FS on $k_xk_y$ and $k_yk_z$ planes are quite similar but different, which are equivalent without SOC. 
FS sheets are labelled as $\alpha$, $\beta$, $\gamma$, and $\delta$ for $\mu=E_F$.
Primes are used to for contours when $\mu=E_F-0.3$ eV.
In the following, aforementioned discussion is presented for $\eta=-5$, $0$, and $+5\%$.
Cases for other strains are shown in Supplemental Material \footnote{See Sec. III of Supplemental Material at http://link.aps.org/supplemental/xxxxx. Fermi surface sheets under different strains are shown for $+1\%$ interval, 
where the Fermi surface contour exhibits smoother and more continuous evolution.}.

%================================================
% Fig.4 volume evolution of BC on plane mu=0
%================================================
%================================================

{\em $\mu=E_F$}:
Contour of FS on $k_xk_y$ and $k_yk_z$ planes are almost the same except for $\gamma$ with $\eta=0\%$.
$\alpha$ is evident for $\eta=-5\%$ and $\eta=0\%$, which disappears for $\eta=+5\%$.
$\beta$ for $\eta=-5\%$ is in arc-shape with nearly two identical bands.
For $\eta=0\%$, $\beta$ is composed of two pieces, denoted as $\beta_1$ and $\beta_2$,
while for $\eta=+5\%$ it is a single piece.
Ellipse-shaped $\delta$ around $K$,  with increasing strain, shifts toward the $\beta$, 
which  intersects $\beta$ for $\eta=+5\%$.
A notable difference between the $k_xk_y$ and $k_yk_z$ planes is prominent in $\gamma$ sheet for $\eta=0\%$.
On $k_xk_y$ plane, $\gamma$ is nearly circular arc shape for $\eta=-5\%$ and elongated for $\eta=0\%$, it disappears for $\eta=+5\%$.
However, on $k_yk_z$ plane, $\gamma$ is closed contour near $M$ for $\eta=0\%$.

While FS on $k_xk_y$ and $k_yk_z$ planes look similar, the Berry curvature distributions differ significantly.  
On $k_xk_y$ plane, $\alpha$ has no appreciable $\Omega$ with and without strain.
For $\eta=-5\%$, $\beta$ has $\Omega>0$ around $k_x$ and $k_y$ axis and $\Omega<0$ around $\frac{1}{2}\Gamma K$.
For $\eta=0\%$, $\beta_1$ and $\beta_2$ have large $\Omega<0$ around $\frac{1}{2}\Gamma K$. 
For $\eta=+5\%$, $\beta$ and $\delta$ have large $\Omega$  around $\frac{1}{2}\Gamma K$.
On $k_yk_z$ plane, $\alpha$ still has no contribution with and without strain.
For $\eta=-5\%$, $\beta$ has both signs of $\Omega$. $\Omega<0$ appears around $k_y$ and $k_z$ axis 
while $\Omega>0$  does near $k_z$ axis.
For $\eta=0\%$, both $\Omega>0$ and $\Omega<0$ emerge near $\frac{1}{2}MX$ 
where nearly degenerate bands are retained and $\Omega<0$ is pronounces in large region outside $\gamma$.
For $\eta=+5\%$, $\beta$ is single-banded with appreciable $\Omega<0$ around crossing with $\delta$ and arc close to $k_z$ axis.

%================================================
% Fig.5 volume evolution of BC on plane mu=-0.3
%================================================
%====================================================
{\em $\mu=E_F-0.3$} eV: The Fermiology, at first sight, is quite similar for all strains %However, the difference lies in detail.
\footnote{See Fig. S7 of Supplemental Material at http://link.aps.org/supplemental/xxxxx. Bands around $E_F-0.3$ eV, near nodal lines do not change with respect to strain.}.
On $k_xk_y$ plane for $\eta=-5\%$, $\alpha'$ is also four-fold but the degeneracy is lifted along $\Gamma-K$.
$\beta'$ is rather straight line further away from $\Gamma$ with respect to $\mu=E_F$.
$\gamma'$ around $X$ is similar to $\mu=E_F$ with smaller size.
$\delta'$ around $K$ is oval-shaped with little distortion.
For $\eta=0\%$, while the overall Fermiology is retained respect to $\eta=-5\%$, the details need some comments.
The lifted degeneracy of $\alpha'$ is larger (smaller) than compressive (tensile) along $\Gamma-K$.
$\gamma'$ is split into $\gamma_1'$ and $\gamma_2'$.
$\delta'$ is smaller.
For $\eta=+5\%$, $\alpha'$ is almost the same as the case for $\eta=0\%$.
$\gamma_1'$, $\gamma_2'$, and $\delta'$ is getting closer and $\gamma_2'$ is collapsed to $\delta'$.
On $k_yk_z$ plane, the Fermiology is remains the same as on $k_xk_y$ plane except for $\eta=+5\%$.
$\gamma_1'$, $\gamma_2'$, and $\delta'$ is getting closer similar to $k_yk_z$, but not collapsed.

Now, we move on to Berry curvature distribution.
On $k_xk_y$ plane, for $\eta=-5\%$, there is no big contribution of $\Omega$.
For $\eta=0$ and $+5\%$, $\Omega>0$ appears along $\Gamma-K$.
On $k_yk_z$ plane, for $\eta=-5\%$, $\Omega<0$ appears near $\alpha'$ and $\gamma'$.
For $\eta=0$ and $+5\%$, $\Omega<0$ contribution are found near $\gamma'_1$, $\gamma'_2$, and $\delta'$, differing in their magnitudes. 
Similar distribution but only the magnitude differences are found for other strains \footnote{See Fig. S5, and S6 of Supplemental Material at
  http://link.aps.org/supplemental/xxxxx. For other strains, 
  the overall Berry curvature distribution is similar with different magnitudes.}.

%====================================================
%.                          CONCLUSION
%====================================================
\section{Conclusion}

We have investigated the anomalous Hall conductivity (AHC) and anomalous Nernst conductivity (ANC) 
of the compensated ferrimagnet Mn$_3$Al under isotropic strain and chemical potential variation, 
based on first-principles calculations. 
When $\mu = -0.3$ eV, three distinct topological features—Weyl points, nodal lines, and gapped nodal lines—are simultaneously realized, occurring along distinct high-symmetry directions in the Brillouin zone in the framework of magnetic space group. 
These symmetry-protected features govern the Berry curvature landscape and strongly influence transverse transport.
Dominant contribution to AHC arises from the $k_yk_z$ plane;
the isotropic strain can significantly enhance AHC, reaching up to $-1200$ $(\Omega~\mathrm{cm})^{-1}$. 
Meanwhile, ANC shows a sign reversal near the Fermi level and 
becomes strongly negative at $\mu = -0.3$ eV for all strains except $\eta=+5\%$.
While the overall band structure and Fermi surface topology are retained under strain, 
the Berry curvature magnitude and distribution evolve markedly.
These results demonstrate how the interlplay of symmetry, topology, and tunable parameters 
such as strain and doping interplay influences transverse transport responses in Mn$_3$Al, 
highlighting its potential as a model system for strain-engineered spintronic applications and Berry curvature-driven transport phenomena.

%==============================================
% Acknowledgement
%===========================================
\begin{acknowledgements}
This research is supported by National Research Foundation (NRF) of Korea
(NRF-RS-2022-NR068225 and  NRF-RS-2024-00449996).
We are also grateful for support by the National Supercomputing Center with supercomputing resources (KSC-2025-CRE-0016) and  computational resource of the UNIST Supercomputing center.
\end{acknowledgements}

%\input{v15.bbl}
%\bibliography{./mn3al_ANE_1}

\begin{thebibliography}{37}%
\makeatletter
\providecommand \@ifxundefined [1]{%
 \@ifx{#1\undefined}
}%
\providecommand \@ifnum [1]{%
 \ifnum #1\expandafter \@firstoftwo
 \else \expandafter \@secondoftwo
 \fi
}%
\providecommand \@ifx [1]{%
 \ifx #1\expandafter \@firstoftwo
 \else \expandafter \@secondoftwo
 \fi
}%
\providecommand \natexlab [1]{#1}%
\providecommand \enquote  [1]{``#1''}%
\providecommand \bibnamefont  [1]{#1}%
\providecommand \bibfnamefont [1]{#1}%
\providecommand \citenamefont [1]{#1}%
\providecommand \href@noop [0]{\@secondoftwo}%
\providecommand \href [0]{\begingroup \@sanitize@url \@href}%
\providecommand \@href[1]{\@@startlink{#1}\@@href}%
\providecommand \@@href[1]{\endgroup#1\@@endlink}%
\providecommand \@sanitize@url [0]{\catcode `\\12\catcode `\$12\catcode
  `\&12\catcode `\#12\catcode `\^12\catcode `\_12\catcode `\%12\relax}%
\providecommand \@@startlink[1]{}%
\providecommand \@@endlink[0]{}%
\providecommand \url  [0]{\begingroup\@sanitize@url \@url }%
\providecommand \@url [1]{\endgroup\@href {#1}{\urlprefix }}%
\providecommand \urlprefix  [0]{URL }%
\providecommand \Eprint [0]{\href }%
\providecommand \doibase [0]{https://doi.org/}%
\providecommand \selectlanguage [0]{\@gobble}%
\providecommand \bibinfo  [0]{\@secondoftwo}%
\providecommand \bibfield  [0]{\@secondoftwo}%
\providecommand \translation [1]{[#1]}%
\providecommand \BibitemOpen [0]{}%
\providecommand \bibitemStop [0]{}%
\providecommand \bibitemNoStop [0]{.\EOS\space}%
\providecommand \EOS [0]{\spacefactor3000\relax}%
\providecommand \BibitemShut  [1]{\csname bibitem#1\endcsname}%
\let\auto@bib@innerbib\@empty
%</preamble>
\bibitem [{\citenamefont {Nakatsuji}\ \emph {et~al.}(2015)\citenamefont
  {Nakatsuji}, \citenamefont {Kiyohara},\ and\ \citenamefont
  {Higo}}]{Nakatsuji:2015aa}%
  \BibitemOpen
  \bibfield  {author} {\bibinfo {author} {\bibfnamefont {S.}~\bibnamefont
  {Nakatsuji}}, \bibinfo {author} {\bibfnamefont {N.}~\bibnamefont
  {Kiyohara}},\ and\ \bibinfo {author} {\bibfnamefont {T.}~\bibnamefont
  {Higo}},\ }\bibfield  {title} {\bibinfo {title} {{Large anomalous Hall effect
  in a non-collinear antiferromagnet at room temperature}},\ }\href
  {https://doi.org/10.1038/nature15723} {\bibfield  {journal} {\bibinfo
  {journal} {Nature}\ }\textbf {\bibinfo {volume} {527}},\ \bibinfo {pages}
  {212} (\bibinfo {year} {2015})}\BibitemShut {NoStop}%
\bibitem [{\citenamefont {Kimata}\ \emph {et~al.}(2019)\citenamefont {Kimata},
  \citenamefont {Chen}, \citenamefont {Kondou}, \citenamefont {Sugimoto},
  \citenamefont {Muduli}, \citenamefont {Ikhlas}, \citenamefont {Omori},
  \citenamefont {Tomita}, \citenamefont {MacDonald}, \citenamefont
  {Nakatsuji},\ and\ \citenamefont {Otani}}]{Kimata:2019aa}%
  \BibitemOpen
  \bibfield  {author} {\bibinfo {author} {\bibfnamefont {M.}~\bibnamefont
  {Kimata}}, \bibinfo {author} {\bibfnamefont {H.}~\bibnamefont {Chen}},
  \bibinfo {author} {\bibfnamefont {K.}~\bibnamefont {Kondou}}, \bibinfo
  {author} {\bibfnamefont {S.}~\bibnamefont {Sugimoto}}, \bibinfo {author}
  {\bibfnamefont {P.~K.}\ \bibnamefont {Muduli}}, \bibinfo {author}
  {\bibfnamefont {M.}~\bibnamefont {Ikhlas}}, \bibinfo {author} {\bibfnamefont
  {Y.}~\bibnamefont {Omori}}, \bibinfo {author} {\bibfnamefont
  {T.}~\bibnamefont {Tomita}}, \bibinfo {author} {\bibfnamefont {A.~H.}\
  \bibnamefont {MacDonald}}, \bibinfo {author} {\bibfnamefont {S.}~\bibnamefont
  {Nakatsuji}},\ and\ \bibinfo {author} {\bibfnamefont {Y.}~\bibnamefont
  {Otani}},\ }\bibfield  {title} {\bibinfo {title} {{Magnetic and magnetic
  inverse spin Hall effects in a non-collinear antiferromagnet}},\ }\href
  {https://doi.org/10.1038/s41586-018-0853-0} {\bibfield  {journal} {\bibinfo
  {journal} {Nature}\ }\textbf {\bibinfo {volume} {565}},\ \bibinfo {pages}
  {627} (\bibinfo {year} {2019})}\BibitemShut {NoStop}%
\bibitem [{\citenamefont {Uchida}\ \emph {et~al.}(2008)\citenamefont {Uchida},
  \citenamefont {Takahashi}, \citenamefont {Harii}, \citenamefont {Ieda},
  \citenamefont {Koshibae}, \citenamefont {Ando}, \citenamefont {Maekawa},\
  and\ \citenamefont {Saitoh}}]{Uchida:nat2008}%
  \BibitemOpen
  \bibfield  {author} {\bibinfo {author} {\bibfnamefont {K.}~\bibnamefont
  {Uchida}}, \bibinfo {author} {\bibfnamefont {S.}~\bibnamefont {Takahashi}},
  \bibinfo {author} {\bibfnamefont {K.}~\bibnamefont {Harii}}, \bibinfo
  {author} {\bibfnamefont {J.}~\bibnamefont {Ieda}}, \bibinfo {author}
  {\bibfnamefont {W.}~\bibnamefont {Koshibae}}, \bibinfo {author}
  {\bibfnamefont {K.}~\bibnamefont {Ando}}, \bibinfo {author} {\bibfnamefont
  {S.}~\bibnamefont {Maekawa}},\ and\ \bibinfo {author} {\bibfnamefont
  {E.}~\bibnamefont {Saitoh}},\ }\bibfield  {title} {\bibinfo {title}
  {{Observation of the spin Seebeck effect}},\ }\href@noop {} {\bibfield
  {journal} {\bibinfo  {journal} {Nature}\ }\textbf {\bibinfo {volume} {455}},\
  \bibinfo {pages} {778} (\bibinfo {year} {2008})}\BibitemShut {NoStop}%
\bibitem [{\citenamefont {Nagaosa}\ \emph {et~al.}(2010)\citenamefont
  {Nagaosa}, \citenamefont {Sinova}, \citenamefont {Onoda}, \citenamefont
  {MacDonald},\ and\ \citenamefont {Ong}}]{Nagaosa_2010}%
  \BibitemOpen
  \bibfield  {author} {\bibinfo {author} {\bibfnamefont {N.}~\bibnamefont
  {Nagaosa}}, \bibinfo {author} {\bibfnamefont {J.}~\bibnamefont {Sinova}},
  \bibinfo {author} {\bibfnamefont {S.}~\bibnamefont {Onoda}}, \bibinfo
  {author} {\bibfnamefont {A.~H.}\ \bibnamefont {MacDonald}},\ and\ \bibinfo
  {author} {\bibfnamefont {N.~P.}\ \bibnamefont {Ong}},\ }\bibfield  {title}
  {\bibinfo {title} {{Anomalous Hall effect}},\ }\href
  {https://doi.org/10.1103/revmodphys.82.1539} {\bibfield  {journal} {\bibinfo
  {journal} {Rev. Mod. Phys.}\ }\textbf {\bibinfo {volume} {82}},\ \bibinfo
  {pages} {1539} (\bibinfo {year} {2010})}\BibitemShut {NoStop}%
\bibitem [{\citenamefont {Chen}\ \emph {et~al.}(2014)\citenamefont {Chen},
  \citenamefont {Niu},\ and\ \citenamefont
  {MacDonald}}]{PhysRevLett.112.017205}%
  \BibitemOpen
  \bibfield  {author} {\bibinfo {author} {\bibfnamefont {H.}~\bibnamefont
  {Chen}}, \bibinfo {author} {\bibfnamefont {Q.}~\bibnamefont {Niu}},\ and\
  \bibinfo {author} {\bibfnamefont {A.~H.}\ \bibnamefont {MacDonald}},\
  }\bibfield  {title} {\bibinfo {title} {{Anomalous Hall Effect Arising from
  Noncollinear Antiferromagnetism}},\ }\href
  {https://doi.org/10.1103/PhysRevLett.112.017205} {\bibfield  {journal}
  {\bibinfo  {journal} {Phys. Rev. Lett.}\ }\textbf {\bibinfo {volume} {112}},\
  \bibinfo {pages} {017205} (\bibinfo {year} {2014})}\BibitemShut {NoStop}%
\bibitem [{\citenamefont {K{\"u}bler}\ and\ \citenamefont
  {Felser}(2014)}]{K_bler_2014}%
  \BibitemOpen
  \bibfield  {author} {\bibinfo {author} {\bibfnamefont {J.}~\bibnamefont
  {K{\"u}bler}}\ and\ \bibinfo {author} {\bibfnamefont {C.}~\bibnamefont
  {Felser}},\ }\bibfield  {title} {\bibinfo {title} {{Non-collinear
  antiferromagnets and the anomalous Hall effect}},\ }\href
  {https://doi.org/10.1209/0295-5075/108/67001} {\bibfield  {journal} {\bibinfo
   {journal} {Europhys. Lett.}\ }\textbf {\bibinfo {volume} {108}},\ \bibinfo
  {pages} {67001} (\bibinfo {year} {2014})}\BibitemShut {NoStop}%
\bibitem [{\citenamefont {Jungwirth}\ \emph {et~al.}(2002)\citenamefont
  {Jungwirth}, \citenamefont {Niu},\ and\ \citenamefont
  {MacDonald}}]{PhysRevLett.88.207208}%
  \BibitemOpen
  \bibfield  {author} {\bibinfo {author} {\bibfnamefont {T.}~\bibnamefont
  {Jungwirth}}, \bibinfo {author} {\bibfnamefont {Q.}~\bibnamefont {Niu}},\
  and\ \bibinfo {author} {\bibfnamefont {A.~H.}\ \bibnamefont {MacDonald}},\
  }\bibfield  {title} {\bibinfo {title} {{Anomalous Hall Effect in
  Ferromagnetic Semiconductors}},\ }\href
  {https://doi.org/10.1103/PhysRevLett.88.207208} {\bibfield  {journal}
  {\bibinfo  {journal} {Phys. Rev. Lett.}\ }\textbf {\bibinfo {volume} {88}},\
  \bibinfo {pages} {207208} (\bibinfo {year} {2002})}\BibitemShut {NoStop}%
\bibitem [{\citenamefont {\ifmmode~\check{S}\else \v{S}\fi{}mejkal}\ \emph
  {et~al.}(2022)\citenamefont {\ifmmode~\check{S}\else \v{S}\fi{}mejkal},
  \citenamefont {Hellenes}, \citenamefont {Gonz\'alez-Hern\'andez},
  \citenamefont {Sinova},\ and\ \citenamefont
  {Jungwirth}}]{PhysRevX.12.011028}%
  \BibitemOpen
  \bibfield  {author} {\bibinfo {author} {\bibfnamefont {L.}~\bibnamefont
  {\ifmmode~\check{S}\else \v{S}\fi{}mejkal}}, \bibinfo {author} {\bibfnamefont
  {A.~B.}\ \bibnamefont {Hellenes}}, \bibinfo {author} {\bibfnamefont
  {R.}~\bibnamefont {Gonz\'alez-Hern\'andez}}, \bibinfo {author} {\bibfnamefont
  {J.}~\bibnamefont {Sinova}},\ and\ \bibinfo {author} {\bibfnamefont
  {T.}~\bibnamefont {Jungwirth}},\ }\bibfield  {title} {\bibinfo {title}
  {{Giant and Tunneling Magnetoresistance in Unconventional Collinear
  Antiferromagnets with Nonrelativistic Spin-Momentum Coupling}},\ }\href
  {https://doi.org/10.1103/PhysRevX.12.011028} {\bibfield  {journal} {\bibinfo
  {journal} {Phys. Rev. X}\ }\textbf {\bibinfo {volume} {12}},\ \bibinfo
  {pages} {011028} (\bibinfo {year} {2022})}\BibitemShut {NoStop}%
\bibitem [{\citenamefont {Xiao}\ \emph {et~al.}(2006)\citenamefont {Xiao},
  \citenamefont {Yao}, \citenamefont {Fang},\ and\ \citenamefont
  {Niu}}]{PhysRevLett.97.026603}%
  \BibitemOpen
  \bibfield  {author} {\bibinfo {author} {\bibfnamefont {D.}~\bibnamefont
  {Xiao}}, \bibinfo {author} {\bibfnamefont {Y.}~\bibnamefont {Yao}}, \bibinfo
  {author} {\bibfnamefont {Z.}~\bibnamefont {Fang}},\ and\ \bibinfo {author}
  {\bibfnamefont {Q.}~\bibnamefont {Niu}},\ }\bibfield  {title} {\bibinfo
  {title} {{Berry-Phase Effect in Anomalous Thermoelectric Transport}},\ }\href
  {https://doi.org/10.1103/PhysRevLett.97.026603} {\bibfield  {journal}
  {\bibinfo  {journal} {Phys. Rev. Lett.}\ }\textbf {\bibinfo {volume} {97}},\
  \bibinfo {pages} {026603} (\bibinfo {year} {2006})}\BibitemShut {NoStop}%
\bibitem [{\citenamefont {Xiao}\ \emph {et~al.}(2010)\citenamefont {Xiao},
  \citenamefont {Chang},\ and\ \citenamefont {Niu}}]{RevModPhys.82.1959}%
  \BibitemOpen
  \bibfield  {author} {\bibinfo {author} {\bibfnamefont {D.}~\bibnamefont
  {Xiao}}, \bibinfo {author} {\bibfnamefont {M.-C.}\ \bibnamefont {Chang}},\
  and\ \bibinfo {author} {\bibfnamefont {Q.}~\bibnamefont {Niu}},\ }\bibfield
  {title} {\bibinfo {title} {{Berry phase effects on electronic properties}},\
  }\href {https://doi.org/10.1103/RevModPhys.82.1959} {\bibfield  {journal}
  {\bibinfo  {journal} {Rev. Mod. Phys.}\ }\textbf {\bibinfo {volume} {82}},\
  \bibinfo {pages} {1959} (\bibinfo {year} {2010})}\BibitemShut {NoStop}%
\bibitem [{\citenamefont {Yao}\ \emph {et~al.}(2004)\citenamefont {Yao},
  \citenamefont {Kleinman}, \citenamefont {MacDonald}, \citenamefont {Sinova},
  \citenamefont {Jungwirth}, \citenamefont {Wang}, \citenamefont {Wang},\ and\
  \citenamefont {Niu}}]{Yao_2004}%
  \BibitemOpen
  \bibfield  {author} {\bibinfo {author} {\bibfnamefont {Y.}~\bibnamefont
  {Yao}}, \bibinfo {author} {\bibfnamefont {L.}~\bibnamefont {Kleinman}},
  \bibinfo {author} {\bibfnamefont {A.~H.}\ \bibnamefont {MacDonald}}, \bibinfo
  {author} {\bibfnamefont {J.}~\bibnamefont {Sinova}}, \bibinfo {author}
  {\bibfnamefont {T.}~\bibnamefont {Jungwirth}}, \bibinfo {author}
  {\bibfnamefont {D.-s.}\ \bibnamefont {Wang}}, \bibinfo {author}
  {\bibfnamefont {E.}~\bibnamefont {Wang}},\ and\ \bibinfo {author}
  {\bibfnamefont {Q.}~\bibnamefont {Niu}},\ }\bibfield  {title} {\bibinfo
  {title} {{First Principles Calculation of Anomalous Hall Conductivity in
  Ferromagnetic bcc Fe}},\ }\href
  {https://doi.org/10.1103/PhysRevLett.92.037204} {\bibfield  {journal}
  {\bibinfo  {journal} {Phys. Rev. Lett.}\ }\textbf {\bibinfo {volume} {92}},\
  \bibinfo {pages} {037204} (\bibinfo {year} {2004})}\BibitemShut {NoStop}%
\bibitem [{\citenamefont {Fang}\ \emph {et~al.}(2003)\citenamefont {Fang},
  \citenamefont {Nagaosa}, \citenamefont {Takahashi}, \citenamefont {Asamitsu},
  \citenamefont {Mathieu}, \citenamefont {Ogasawara}, \citenamefont {Yamada},
  \citenamefont {Kawasaki}, \citenamefont {Tokura},\ and\ \citenamefont
  {Terakura}}]{ZFang:Sci-2003}%
  \BibitemOpen
  \bibfield  {author} {\bibinfo {author} {\bibfnamefont {Z.}~\bibnamefont
  {Fang}}, \bibinfo {author} {\bibfnamefont {N.}~\bibnamefont {Nagaosa}},
  \bibinfo {author} {\bibfnamefont {K.~S.}\ \bibnamefont {Takahashi}}, \bibinfo
  {author} {\bibfnamefont {A.}~\bibnamefont {Asamitsu}}, \bibinfo {author}
  {\bibfnamefont {R.}~\bibnamefont {Mathieu}}, \bibinfo {author} {\bibfnamefont
  {T.}~\bibnamefont {Ogasawara}}, \bibinfo {author} {\bibfnamefont
  {H.}~\bibnamefont {Yamada}}, \bibinfo {author} {\bibfnamefont
  {M.}~\bibnamefont {Kawasaki}}, \bibinfo {author} {\bibfnamefont
  {Y.}~\bibnamefont {Tokura}},\ and\ \bibinfo {author} {\bibfnamefont
  {K.}~\bibnamefont {Terakura}},\ }\bibfield  {title} {\bibinfo {title} {{The
  Anomalous Hall Effect and Magnetic Monopoles in Momentum Space}},\
  }\href@noop {} {\bibfield  {journal} {\bibinfo  {journal} {Science}\ }\textbf
  {\bibinfo {volume} {302}},\ \bibinfo {pages} {92} (\bibinfo {year}
  {2003})}\BibitemShut {NoStop}%
\bibitem [{\citenamefont {{\v S}mejkal}\ \emph {et~al.}(2022)\citenamefont {{\v
  S}mejkal}, \citenamefont {MacDonald}, \citenamefont {Sinova}, \citenamefont
  {Nakatsuji},\ and\ \citenamefont {Jungwirth}}]{Smejkal:2022aa}%
  \BibitemOpen
  \bibfield  {author} {\bibinfo {author} {\bibfnamefont {L.}~\bibnamefont {{\v
  S}mejkal}}, \bibinfo {author} {\bibfnamefont {A.~H.}\ \bibnamefont
  {MacDonald}}, \bibinfo {author} {\bibfnamefont {J.}~\bibnamefont {Sinova}},
  \bibinfo {author} {\bibfnamefont {S.}~\bibnamefont {Nakatsuji}},\ and\
  \bibinfo {author} {\bibfnamefont {T.}~\bibnamefont {Jungwirth}},\ }\bibfield
  {title} {\bibinfo {title} {{Anomalous Hall antiferromagnets}},\ }\href
  {https://doi.org/10.1038/s41578-022-00430-3} {\bibfield  {journal} {\bibinfo
  {journal} {Nat. Rev. Mater.}\ }\textbf {\bibinfo {volume} {7}},\ \bibinfo
  {pages} {482} (\bibinfo {year} {2022})}\BibitemShut {NoStop}%
\bibitem [{\citenamefont {Baltz}\ \emph {et~al.}(2018)\citenamefont {Baltz},
  \citenamefont {Manchon}, \citenamefont {Tsoi}, \citenamefont {Moriyama},
  \citenamefont {Ono},\ and\ \citenamefont
  {Tserkovnyak}}]{RevModPhys.90.015005}%
  \BibitemOpen
  \bibfield  {author} {\bibinfo {author} {\bibfnamefont {V.}~\bibnamefont
  {Baltz}}, \bibinfo {author} {\bibfnamefont {A.}~\bibnamefont {Manchon}},
  \bibinfo {author} {\bibfnamefont {M.}~\bibnamefont {Tsoi}}, \bibinfo {author}
  {\bibfnamefont {T.}~\bibnamefont {Moriyama}}, \bibinfo {author}
  {\bibfnamefont {T.}~\bibnamefont {Ono}},\ and\ \bibinfo {author}
  {\bibfnamefont {Y.}~\bibnamefont {Tserkovnyak}},\ }\bibfield  {title}
  {\bibinfo {title} {Antiferromagnetic spintronics},\ }\href
  {https://doi.org/10.1103/RevModPhys.90.015005} {\bibfield  {journal}
  {\bibinfo  {journal} {Rev. Mod. Phys.}\ }\textbf {\bibinfo {volume} {90}},\
  \bibinfo {pages} {015005} (\bibinfo {year} {2018})}\BibitemShut {NoStop}%
\bibitem [{\citenamefont {Jungwirth}\ \emph {et~al.}(2016)\citenamefont
  {Jungwirth}, \citenamefont {Marti}, \citenamefont {Wadley},\ and\
  \citenamefont {Wunderlich}}]{Jungwirth:2016aa}%
  \BibitemOpen
  \bibfield  {author} {\bibinfo {author} {\bibfnamefont {T.}~\bibnamefont
  {Jungwirth}}, \bibinfo {author} {\bibfnamefont {X.}~\bibnamefont {Marti}},
  \bibinfo {author} {\bibfnamefont {P.}~\bibnamefont {Wadley}},\ and\ \bibinfo
  {author} {\bibfnamefont {J.}~\bibnamefont {Wunderlich}},\ }\bibfield  {title}
  {\bibinfo {title} {{Antiferromagnetic spintronics}},\ }\href
  {https://doi.org/10.1038/nnano.2016.18} {\bibfield  {journal} {\bibinfo
  {journal} {Nat. Nanotechnol.}\ }\textbf {\bibinfo {volume} {11}},\ \bibinfo
  {pages} {231} (\bibinfo {year} {2016})}\BibitemShut {NoStop}%
\bibitem [{\citenamefont {Graf}\ \emph {et~al.}(2011)\citenamefont {Graf},
  \citenamefont {Felser},\ and\ \citenamefont {Parkin}}]{GRAF20111}%
  \BibitemOpen
  \bibfield  {author} {\bibinfo {author} {\bibfnamefont {T.}~\bibnamefont
  {Graf}}, \bibinfo {author} {\bibfnamefont {C.}~\bibnamefont {Felser}},\ and\
  \bibinfo {author} {\bibfnamefont {S.~S.}\ \bibnamefont {Parkin}},\ }\bibfield
   {title} {\bibinfo {title} {{Simple rules for the understanding of Heusler
  compounds}},\ }\href
  {https://doi.org/https://doi.org/10.1016/j.progsolidstchem.2011.02.001}
  {\bibfield  {journal} {\bibinfo  {journal} {Prog. Solid State Chem.}\
  }\textbf {\bibinfo {volume} {39}},\ \bibinfo {pages} {1} (\bibinfo {year}
  {2011})}\BibitemShut {NoStop}%
\bibitem [{\citenamefont {K\"ubler}\ and\ \citenamefont
  {Felser}(2012)}]{PhysRevB.85.012405}%
  \BibitemOpen
  \bibfield  {author} {\bibinfo {author} {\bibfnamefont {J.}~\bibnamefont
  {K\"ubler}}\ and\ \bibinfo {author} {\bibfnamefont {C.}~\bibnamefont
  {Felser}},\ }\bibfield  {title} {\bibinfo {title} {{Berry curvature and the
  anomalous Hall effect in Heusler compounds}},\ }\href
  {https://doi.org/10.1103/PhysRevB.85.012405} {\bibfield  {journal} {\bibinfo
  {journal} {Phys. Rev. B}\ }\textbf {\bibinfo {volume} {85}},\ \bibinfo
  {pages} {012405} (\bibinfo {year} {2012})}\BibitemShut {NoStop}%
\bibitem [{\citenamefont {Wollmann}\ \emph {et~al.}(2017)\citenamefont
  {Wollmann}, \citenamefont {Nayak}, \citenamefont {Parkin},\ and\
  \citenamefont {Felser}}]{Wollmann-ann.rev.mat:2017}%
  \BibitemOpen
  \bibfield  {author} {\bibinfo {author} {\bibfnamefont {L.}~\bibnamefont
  {Wollmann}}, \bibinfo {author} {\bibfnamefont {A.~K.}\ \bibnamefont {Nayak}},
  \bibinfo {author} {\bibfnamefont {S.~S.}\ \bibnamefont {Parkin}},\ and\
  \bibinfo {author} {\bibfnamefont {C.}~\bibnamefont {Felser}},\ }\bibfield
  {title} {\bibinfo {title} {Heusler 4.0: Tunable materials},\ }\href
  {https://doi.org/https://doi.org/10.1146/annurev-matsci-070616-123928}
  {\bibfield  {journal} {\bibinfo  {journal} {Annu. Rev. Mater. Res}\ }\textbf
  {\bibinfo {volume} {47}},\ \bibinfo {pages} {247} (\bibinfo {year}
  {2017})}\BibitemShut {NoStop}%
\bibitem [{\citenamefont {Jamer}\ \emph {et~al.}(2017)\citenamefont {Jamer},
  \citenamefont {Wang}, \citenamefont {Stephen}, \citenamefont {McDonald},
  \citenamefont {Grutter}, \citenamefont {Sterbinsky}, \citenamefont {Arena},
  \citenamefont {Borchers}, \citenamefont {Kirby}, \citenamefont {Lewis},
  \citenamefont {Barbiellini}, \citenamefont {Bansil},\ and\ \citenamefont
  {Heiman}}]{Jamer_2017}%
  \BibitemOpen
  \bibfield  {author} {\bibinfo {author} {\bibfnamefont {M.~E.}\ \bibnamefont
  {Jamer}}, \bibinfo {author} {\bibfnamefont {Y.~J.}\ \bibnamefont {Wang}},
  \bibinfo {author} {\bibfnamefont {G.~M.}\ \bibnamefont {Stephen}}, \bibinfo
  {author} {\bibfnamefont {I.~J.}\ \bibnamefont {McDonald}}, \bibinfo {author}
  {\bibfnamefont {A.~J.}\ \bibnamefont {Grutter}}, \bibinfo {author}
  {\bibfnamefont {G.~E.}\ \bibnamefont {Sterbinsky}}, \bibinfo {author}
  {\bibfnamefont {D.~A.}\ \bibnamefont {Arena}}, \bibinfo {author}
  {\bibfnamefont {J.~A.}\ \bibnamefont {Borchers}}, \bibinfo {author}
  {\bibfnamefont {B.~J.}\ \bibnamefont {Kirby}}, \bibinfo {author}
  {\bibfnamefont {L.~H.}\ \bibnamefont {Lewis}}, \bibinfo {author}
  {\bibfnamefont {B.}~\bibnamefont {Barbiellini}}, \bibinfo {author}
  {\bibfnamefont {A.}~\bibnamefont {Bansil}},\ and\ \bibinfo {author}
  {\bibfnamefont {D.}~\bibnamefont {Heiman}},\ }\bibfield  {title} {\bibinfo
  {title} {{Compensated Ferrimagnetism in the Zero-Moment Heusler Alloy
  ${\mathrm{Mn}}_{3}\mathrm{Al}$}},\ }\href
  {https://doi.org/10.1103/PhysRevApplied.7.064036} {\bibfield  {journal}
  {\bibinfo  {journal} {Phys. Rev. Appl.}\ }\textbf {\bibinfo {volume} {7}},\
  \bibinfo {pages} {064036} (\bibinfo {year} {2017})}\BibitemShut {NoStop}%
\bibitem [{\citenamefont {Park}\ \emph {et~al.}(2022)\citenamefont {Park},
  \citenamefont {Han},\ and\ \citenamefont {Rhim}}]{PhysRevResearch.4.013215}%
  \BibitemOpen
  \bibfield  {author} {\bibinfo {author} {\bibfnamefont {M.}~\bibnamefont
  {Park}}, \bibinfo {author} {\bibfnamefont {G.}~\bibnamefont {Han}},\ and\
  \bibinfo {author} {\bibfnamefont {S.~H.}\ \bibnamefont {Rhim}},\ }\bibfield
  {title} {\bibinfo {title} {{Anomalous Hall effect in a compensated
  ferrimagnet: Symmetry analysis for ${\mathrm{Mn}}_{3}\mathrm{Al}$}},\ }\href
  {https://doi.org/10.1103/PhysRevResearch.4.013215} {\bibfield  {journal}
  {\bibinfo  {journal} {Phys. Rev. Res.}\ }\textbf {\bibinfo {volume} {4}},\
  \bibinfo {pages} {013215} (\bibinfo {year} {2022})}\BibitemShut {NoStop}%
\bibitem [{\citenamefont {Shi}\ \emph {et~al.}(2018)\citenamefont {Shi},
  \citenamefont {Muechler}, \citenamefont {Manna}, \citenamefont {Zhang},
  \citenamefont {Koepernik}, \citenamefont {Car}, \citenamefont {van~den
  Brink}, \citenamefont {Felser},\ and\ \citenamefont
  {Sun}}]{PhysRevB.97.060406}%
  \BibitemOpen
  \bibfield  {author} {\bibinfo {author} {\bibfnamefont {W.}~\bibnamefont
  {Shi}}, \bibinfo {author} {\bibfnamefont {L.}~\bibnamefont {Muechler}},
  \bibinfo {author} {\bibfnamefont {K.}~\bibnamefont {Manna}}, \bibinfo
  {author} {\bibfnamefont {Y.}~\bibnamefont {Zhang}}, \bibinfo {author}
  {\bibfnamefont {K.}~\bibnamefont {Koepernik}}, \bibinfo {author}
  {\bibfnamefont {R.}~\bibnamefont {Car}}, \bibinfo {author} {\bibfnamefont
  {J.}~\bibnamefont {van~den Brink}}, \bibinfo {author} {\bibfnamefont
  {C.}~\bibnamefont {Felser}},\ and\ \bibinfo {author} {\bibfnamefont
  {Y.}~\bibnamefont {Sun}},\ }\bibfield  {title} {\bibinfo {title} {{Prediction
  of a magnetic Weyl semimetal without spin-orbit coupling and strong anomalous
  Hall effect in the Heusler compensated ferrimagnet
  ${\mathrm{Ti}}_{2}\mathrm{MnAl}$}},\ }\href
  {https://doi.org/10.1103/PhysRevB.97.060406} {\bibfield  {journal} {\bibinfo
  {journal} {Phys. Rev. B}\ }\textbf {\bibinfo {volume} {97}},\ \bibinfo
  {pages} {060406} (\bibinfo {year} {2018})}\BibitemShut {NoStop}%
\bibitem [{\citenamefont {Zhang}\ \emph {et~al.}(2017)\citenamefont {Zhang},
  \citenamefont {Sun}, \citenamefont {Yang}, \citenamefont
  {\ifmmode~\check{Z}\else \v{Z}\fi{}elezn\'y}, \citenamefont {Parkin},
  \citenamefont {Felser},\ and\ \citenamefont {Yan}}]{PhysRevB.95.075128}%
  \BibitemOpen
  \bibfield  {author} {\bibinfo {author} {\bibfnamefont {Y.}~\bibnamefont
  {Zhang}}, \bibinfo {author} {\bibfnamefont {Y.}~\bibnamefont {Sun}}, \bibinfo
  {author} {\bibfnamefont {H.}~\bibnamefont {Yang}}, \bibinfo {author}
  {\bibfnamefont {J.}~\bibnamefont {\ifmmode~\check{Z}\else
  \v{Z}\fi{}elezn\'y}}, \bibinfo {author} {\bibfnamefont {S.~P.~P.}\
  \bibnamefont {Parkin}}, \bibinfo {author} {\bibfnamefont {C.}~\bibnamefont
  {Felser}},\ and\ \bibinfo {author} {\bibfnamefont {B.}~\bibnamefont {Yan}},\
  }\bibfield  {title} {\bibinfo {title} {{Strong anisotropic anomalous Hall
  effect and spin Hall effect in the chiral antiferromagnetic compounds
  ${\mathrm{Mn}}_{3}X$ ($X=\mathrm{Ge}$, Sn, Ga, Ir, Rh, and Pt)}},\ }\href
  {https://doi.org/10.1103/PhysRevB.95.075128} {\bibfield  {journal} {\bibinfo
  {journal} {Phys. Rev. B}\ }\textbf {\bibinfo {volume} {95}},\ \bibinfo
  {pages} {075128} (\bibinfo {year} {2017})}\BibitemShut {NoStop}%
\bibitem [{\citenamefont {{\v S}mejkal}\ \emph {et~al.}(2020)\citenamefont {{\v
  S}mejkal}, \citenamefont {Gonz{\'a}lez-Hern{\'a}ndez}, \citenamefont
  {Jungwirth},\ and\ \citenamefont {Sinova}}]{Smejkal:2020fg}%
  \BibitemOpen
  \bibfield  {author} {\bibinfo {author} {\bibfnamefont {L.}~\bibnamefont {{\v
  S}mejkal}}, \bibinfo {author} {\bibfnamefont {R.}~\bibnamefont
  {Gonz{\'a}lez-Hern{\'a}ndez}}, \bibinfo {author} {\bibfnamefont
  {T.}~\bibnamefont {Jungwirth}},\ and\ \bibinfo {author} {\bibfnamefont
  {J.}~\bibnamefont {Sinova}},\ }\bibfield  {title} {\bibinfo {title} {{Crystal
  time-reversal symmetry breaking and spontaneous Hall effect in collinear
  antiferromagnets}},\ }\href {https://doi.org/10.1126/sciadv.aaz8809}
  {\bibfield  {journal} {\bibinfo  {journal} {Sci. Adv.}\ }\textbf {\bibinfo
  {volume} {6}},\ \bibinfo {pages} {eaaz8809} (\bibinfo {year}
  {2020})}\BibitemShut {NoStop}%
\bibitem [{\citenamefont {Noky}\ \emph {et~al.}(2019)\citenamefont {Noky},
  \citenamefont {Xu}, \citenamefont {Felser},\ and\ \citenamefont
  {Sun}}]{PhysRevB.99.165117}%
  \BibitemOpen
  \bibfield  {author} {\bibinfo {author} {\bibfnamefont {J.}~\bibnamefont
  {Noky}}, \bibinfo {author} {\bibfnamefont {Q.}~\bibnamefont {Xu}}, \bibinfo
  {author} {\bibfnamefont {C.}~\bibnamefont {Felser}},\ and\ \bibinfo {author}
  {\bibfnamefont {Y.}~\bibnamefont {Sun}},\ }\bibfield  {title} {\bibinfo
  {title} {{Large anomalous Hall and Nernst effects from nodal line symmetry
  breaking in ${\mathrm{Fe}}_{2}\mathrm{Mn}X$ ($X$ = P, As, Sb)}},\ }\href
  {https://doi.org/10.1103/PhysRevB.99.165117} {\bibfield  {journal} {\bibinfo
  {journal} {Phys. Rev. B}\ }\textbf {\bibinfo {volume} {99}},\ \bibinfo
  {pages} {165117} (\bibinfo {year} {2019})}\BibitemShut {NoStop}%
\bibitem [{\citenamefont {Li}\ \emph {et~al.}(2020)\citenamefont {Li},
  \citenamefont {Koo}, \citenamefont {Ning}, \citenamefont {Li}, \citenamefont
  {Miao}, \citenamefont {Min}, \citenamefont {Zhu}, \citenamefont {Wang},
  \citenamefont {Alem}, \citenamefont {Liu}, \citenamefont {Mao},\ and\
  \citenamefont {Yan}}]{Li:2020uj}%
  \BibitemOpen
  \bibfield  {author} {\bibinfo {author} {\bibfnamefont {P.}~\bibnamefont
  {Li}}, \bibinfo {author} {\bibfnamefont {J.}~\bibnamefont {Koo}}, \bibinfo
  {author} {\bibfnamefont {W.}~\bibnamefont {Ning}}, \bibinfo {author}
  {\bibfnamefont {J.}~\bibnamefont {Li}}, \bibinfo {author} {\bibfnamefont
  {L.}~\bibnamefont {Miao}}, \bibinfo {author} {\bibfnamefont {L.}~\bibnamefont
  {Min}}, \bibinfo {author} {\bibfnamefont {Y.}~\bibnamefont {Zhu}}, \bibinfo
  {author} {\bibfnamefont {Y.}~\bibnamefont {Wang}}, \bibinfo {author}
  {\bibfnamefont {N.}~\bibnamefont {Alem}}, \bibinfo {author} {\bibfnamefont
  {C.-X.}\ \bibnamefont {Liu}}, \bibinfo {author} {\bibfnamefont
  {Z.}~\bibnamefont {Mao}},\ and\ \bibinfo {author} {\bibfnamefont
  {B.}~\bibnamefont {Yan}},\ }\bibfield  {title} {\bibinfo {title} {{Giant room
  temperature anomalous Hall effect and tunable topology in a ferromagnetic
  topological semimetal Co$_2$MnAl}},\ }\href
  {https://doi.org/10.1038/s41467-020-17174-9} {\bibfield  {journal} {\bibinfo
  {journal} {Nat. Commun.}\ }\textbf {\bibinfo {volume} {11}},\ \bibinfo
  {pages} {3476} (\bibinfo {year} {2020})}\BibitemShut {NoStop}%
\bibitem [{\citenamefont {Sun}\ \emph {et~al.}(2021)\citenamefont {Sun},
  \citenamefont {Peng}, \citenamefont {Cui}, \citenamefont {Zhu}, \citenamefont
  {Zhuo}, \citenamefont {Wang},\ and\ \citenamefont
  {Chen}}]{PhysRevB.103.085116}%
  \BibitemOpen
  \bibfield  {author} {\bibinfo {author} {\bibfnamefont {Z.~L.}\ \bibnamefont
  {Sun}}, \bibinfo {author} {\bibfnamefont {K.~L.}\ \bibnamefont {Peng}},
  \bibinfo {author} {\bibfnamefont {J.~H.}\ \bibnamefont {Cui}}, \bibinfo
  {author} {\bibfnamefont {C.~S.}\ \bibnamefont {Zhu}}, \bibinfo {author}
  {\bibfnamefont {W.~Z.}\ \bibnamefont {Zhuo}}, \bibinfo {author}
  {\bibfnamefont {Z.~Y.}\ \bibnamefont {Wang}},\ and\ \bibinfo {author}
  {\bibfnamefont {X.~H.}\ \bibnamefont {Chen}},\ }\bibfield  {title} {\bibinfo
  {title} {{Pressure-controlled anomalous Hall conductivity in the half-Heusler
  antiferromagnet GdPtBi}},\ }\href
  {https://doi.org/10.1103/PhysRevB.103.085116} {\bibfield  {journal} {\bibinfo
   {journal} {Phys. Rev. B}\ }\textbf {\bibinfo {volume} {103}},\ \bibinfo
  {pages} {085116} (\bibinfo {year} {2021})}\BibitemShut {NoStop}%
\bibitem [{\citenamefont {Singh}\ \emph {et~al.}(2020)\citenamefont {Singh},
  \citenamefont {Singh}, \citenamefont {Pradhan}, \citenamefont {Srihari},
  \citenamefont {Poswal}, \citenamefont {Nath}, \citenamefont {Nandy},\ and\
  \citenamefont {Nayak}}]{PhysRevResearch.2.043366}%
  \BibitemOpen
  \bibfield  {author} {\bibinfo {author} {\bibfnamefont {C.}~\bibnamefont
  {Singh}}, \bibinfo {author} {\bibfnamefont {V.}~\bibnamefont {Singh}},
  \bibinfo {author} {\bibfnamefont {G.}~\bibnamefont {Pradhan}}, \bibinfo
  {author} {\bibfnamefont {V.}~\bibnamefont {Srihari}}, \bibinfo {author}
  {\bibfnamefont {H.~K.}\ \bibnamefont {Poswal}}, \bibinfo {author}
  {\bibfnamefont {R.}~\bibnamefont {Nath}}, \bibinfo {author} {\bibfnamefont
  {A.~K.}\ \bibnamefont {Nandy}},\ and\ \bibinfo {author} {\bibfnamefont
  {A.~K.}\ \bibnamefont {Nayak}},\ }\bibfield  {title} {\bibinfo {title}
  {{Pressure controlled trimerization for switching of anomalous Hall effect in
  triangular antiferromagnet ${\mathrm{Mn}}_{3}\mathrm{Sn}$}},\ }\href
  {https://doi.org/10.1103/PhysRevResearch.2.043366} {\bibfield  {journal}
  {\bibinfo  {journal} {Phys. Rev. Res.}\ }\textbf {\bibinfo {volume} {2}},\
  \bibinfo {pages} {043366} (\bibinfo {year} {2020})}\BibitemShut {NoStop}%
\bibitem [{\citenamefont {Reis}\ \emph {et~al.}(2020)\citenamefont {Reis},
  \citenamefont {Ghorbani~Zavareh}, \citenamefont {Ajeesh}, \citenamefont
  {Kutelak}, \citenamefont {Sukhanov}, \citenamefont {Singh}, \citenamefont
  {Noky}, \citenamefont {Sun}, \citenamefont {Fischer}, \citenamefont {Manna},
  \citenamefont {Felser},\ and\ \citenamefont
  {Nicklas}}]{PhysRevMaterials.4.051401}%
  \BibitemOpen
  \bibfield  {author} {\bibinfo {author} {\bibfnamefont {R.~D.~d.}\
  \bibnamefont {Reis}}, \bibinfo {author} {\bibfnamefont {M.}~\bibnamefont
  {Ghorbani~Zavareh}}, \bibinfo {author} {\bibfnamefont {M.~O.}\ \bibnamefont
  {Ajeesh}}, \bibinfo {author} {\bibfnamefont {L.~O.}\ \bibnamefont {Kutelak}},
  \bibinfo {author} {\bibfnamefont {A.~S.}\ \bibnamefont {Sukhanov}}, \bibinfo
  {author} {\bibfnamefont {S.}~\bibnamefont {Singh}}, \bibinfo {author}
  {\bibfnamefont {J.}~\bibnamefont {Noky}}, \bibinfo {author} {\bibfnamefont
  {Y.}~\bibnamefont {Sun}}, \bibinfo {author} {\bibfnamefont {J.~E.}\
  \bibnamefont {Fischer}}, \bibinfo {author} {\bibfnamefont {K.}~\bibnamefont
  {Manna}}, \bibinfo {author} {\bibfnamefont {C.}~\bibnamefont {Felser}},\ and\
  \bibinfo {author} {\bibfnamefont {M.}~\bibnamefont {Nicklas}},\ }\bibfield
  {title} {\bibinfo {title} {{Pressure tuning of the anomalous Hall effect in
  the chiral antiferromagnet ${\mathrm{Mn}}_{3}\mathrm{Ge}$}},\ }\href
  {https://doi.org/10.1103/PhysRevMaterials.4.051401} {\bibfield  {journal}
  {\bibinfo  {journal} {Phys. Rev. Mater.}\ }\textbf {\bibinfo {volume} {4}},\
  \bibinfo {pages} {051401} (\bibinfo {year} {2020})}\BibitemShut {NoStop}%
  \bibitem [{\citenamefont {Sukhanov}\ \emph {et~al.}(2018)\citenamefont
  {Sukhanov}, \citenamefont {Singh}, \citenamefont {Caron}, \citenamefont
  {Hansen}, \citenamefont {Hoser}, \citenamefont {Kumar}, \citenamefont
  {Borrmann}, \citenamefont {Fitch}, \citenamefont {Devi}, \citenamefont
  {Manna}, \citenamefont {Felser},\ and\ \citenamefont
  {Inosov}}]{PhysRevB.97.214402}%
  \BibitemOpen
  \bibfield  {author} {\bibinfo {author} {\bibfnamefont {A.~S.}\ \bibnamefont
  {Sukhanov}}, \bibinfo {author} {\bibfnamefont {S.}~\bibnamefont {Singh}},
  \bibinfo {author} {\bibfnamefont {L.}~\bibnamefont {Caron}}, \bibinfo
  {author} {\bibfnamefont {T.}~\bibnamefont {Hansen}}, \bibinfo {author}
  {\bibfnamefont {A.}~\bibnamefont {Hoser}}, \bibinfo {author} {\bibfnamefont
  {V.}~\bibnamefont {Kumar}}, \bibinfo {author} {\bibfnamefont
  {H.}~\bibnamefont {Borrmann}}, \bibinfo {author} {\bibfnamefont
  {A.}~\bibnamefont {Fitch}}, \bibinfo {author} {\bibfnamefont
  {P.}~\bibnamefont {Devi}}, \bibinfo {author} {\bibfnamefont {K.}~\bibnamefont
  {Manna}}, \bibinfo {author} {\bibfnamefont {C.}~\bibnamefont {Felser}},\ and\
  \bibinfo {author} {\bibfnamefont {D.~S.}\ \bibnamefont {Inosov}},\ }\bibfield
   {title} {\bibinfo {title} {{Gradual pressure-induced change in the magnetic
  structure of the noncollinear antiferromagnet
  ${\mathrm{Mn}}_{3}\mathrm{Ge}$}},\ }\href
  {https://doi.org/10.1103/PhysRevB.97.214402} {\bibfield  {journal} {\bibinfo
  {journal} {Phys. Rev. B}\ }\textbf {\bibinfo {volume} {97}},\ \bibinfo
  {pages} {214402} (\bibinfo {year} {2018})}\BibitemShut {NoStop}%
\bibitem [{\citenamefont {Kim}\ \emph {et~al.}(2022)\citenamefont {Kim},
  \citenamefont {Pathak}, \citenamefont {Rhim}, \citenamefont {Cha},
  \citenamefont {Jekal}, \citenamefont {Hong}, \citenamefont {Lee},
  \citenamefont {Park}, \citenamefont {Lee}, \citenamefont {Park},
  \citenamefont {Lee}, \citenamefont {Steinr{\"u}ck}, \citenamefont {Mehta},
  \citenamefont {Wang},\ and\ \citenamefont
  {Hong}}]{https://doi.org/10.1002/advs.202201749}%
  \BibitemOpen
  \bibfield  {author} {\bibinfo {author} {\bibfnamefont {S.}~\bibnamefont
  {Kim}}, \bibinfo {author} {\bibfnamefont {S.}~\bibnamefont {Pathak}},
  \bibinfo {author} {\bibfnamefont {S.~H.}\ \bibnamefont {Rhim}}, \bibinfo
  {author} {\bibfnamefont {J.}~\bibnamefont {Cha}}, \bibinfo {author}
  {\bibfnamefont {S.}~\bibnamefont {Jekal}}, \bibinfo {author} {\bibfnamefont
  {S.~C.}\ \bibnamefont {Hong}}, \bibinfo {author} {\bibfnamefont {H.~H.}\
  \bibnamefont {Lee}}, \bibinfo {author} {\bibfnamefont {S.-H.}\ \bibnamefont
  {Park}}, \bibinfo {author} {\bibfnamefont {H.-K.}\ \bibnamefont {Lee}},
  \bibinfo {author} {\bibfnamefont {J.-H.}\ \bibnamefont {Park}}, \bibinfo
  {author} {\bibfnamefont {S.}~\bibnamefont {Lee}}, \bibinfo {author}
  {\bibfnamefont {H.-G.}\ \bibnamefont {Steinr{\"u}ck}}, \bibinfo {author}
  {\bibfnamefont {A.}~\bibnamefont {Mehta}}, \bibinfo {author} {\bibfnamefont
  {S.~X.}\ \bibnamefont {Wang}},\ and\ \bibinfo {author} {\bibfnamefont
  {J.}~\bibnamefont {Hong}},\ }\bibfield  {title} {\bibinfo {title} {{Giant
  Orbital Anisotropy with Strong Spin--Orbit Coupling Established at the
  Pseudomorphic Interface of the Co/Pd Superlattice}},\ }\href
  {https://doi.org/https://doi.org/10.1002/advs.202201749} {\bibfield
  {journal} {\bibinfo  {journal} {Adv. Sci.}\ }\textbf {\bibinfo {volume}
  {9}},\ \bibinfo {pages} {2201749} (\bibinfo {year} {2022})}\BibitemShut
  {NoStop}%
\bibitem [{\citenamefont {Son}\ \emph {et~al.}(2016)\citenamefont {Son},
  \citenamefont {Lee}, \citenamefont {Kim}, \citenamefont {Park}, \citenamefont
  {Lee}, \citenamefont {Kim}, \citenamefont {Kim}, \citenamefont {Hong},\ and\
  \citenamefont {Hong}}]{Son:2016aa}%
  \BibitemOpen
  \bibfield  {author} {\bibinfo {author} {\bibfnamefont {J.}~\bibnamefont
  {Son}}, \bibinfo {author} {\bibfnamefont {S.}~\bibnamefont {Lee}}, \bibinfo
  {author} {\bibfnamefont {S.~J.}\ \bibnamefont {Kim}}, \bibinfo {author}
  {\bibfnamefont {B.~C.}\ \bibnamefont {Park}}, \bibinfo {author}
  {\bibfnamefont {H.-K.}\ \bibnamefont {Lee}}, \bibinfo {author} {\bibfnamefont
  {S.}~\bibnamefont {Kim}}, \bibinfo {author} {\bibfnamefont {J.~H.}\
  \bibnamefont {Kim}}, \bibinfo {author} {\bibfnamefont {B.~H.}\ \bibnamefont
  {Hong}},\ and\ \bibinfo {author} {\bibfnamefont {J.}~\bibnamefont {Hong}},\
  }\bibfield  {title} {\bibinfo {title} {{Hydrogenated monolayer graphene with
  reversible and tunable wide band gap and its field-effect transistor}},\
  }\href {https://doi.org/10.1038/ncomms13261} {\bibfield  {journal} {\bibinfo
  {journal} {Nat. Commun.}\ }\textbf {\bibinfo {volume} {7}},\ \bibinfo {pages}
  {13261} (\bibinfo {year} {2016})}\BibitemShut {NoStop}%
\bibitem [{\citenamefont {Kresse}\ and\ \citenamefont
  {Furthm\"uller}(1996)}]{vasp}%
  \BibitemOpen
  \bibfield  {author} {\bibinfo {author} {\bibfnamefont {G.}~\bibnamefont
  {Kresse}}\ and\ \bibinfo {author} {\bibfnamefont {J.}~\bibnamefont
  {Furthm\"uller}},\ }\bibfield  {title} {\bibinfo {title} {{Efficient
  iterative schemes for ab initio total-energy calculations using a plane-wave
  basis set}},\ }\href {https://doi.org/10.1103/PhysRevB.54.11169} {\bibfield
  {journal} {\bibinfo  {journal} {Phys. Rev. B}\ }\textbf {\bibinfo {volume}
  {54}},\ \bibinfo {pages} {11169} (\bibinfo {year} {1996})}\BibitemShut
  {NoStop}%
\bibitem [{\citenamefont {Perdew}\ \emph {et~al.}(1996)\citenamefont {Perdew},
  \citenamefont {Burke},\ and\ \citenamefont
  {Ernzerhof}}]{PhysRevLett.77.3865}%
  \BibitemOpen
  \bibfield  {author} {\bibinfo {author} {\bibfnamefont {J.~P.}\ \bibnamefont
  {Perdew}}, \bibinfo {author} {\bibfnamefont {K.}~\bibnamefont {Burke}},\ and\
  \bibinfo {author} {\bibfnamefont {M.}~\bibnamefont {Ernzerhof}},\ }\bibfield
  {title} {\bibinfo {title} {{Generalized Gradient Approximation Made
  Simple}},\ }\href {https://doi.org/10.1103/PhysRevLett.77.3865} {\bibfield
  {journal} {\bibinfo  {journal} {Phys. Rev. Lett.}\ }\textbf {\bibinfo
  {volume} {77}},\ \bibinfo {pages} {3865} (\bibinfo {year}
  {1996})}\BibitemShut {NoStop}%
\bibitem [{\citenamefont {Wang}\ \emph {et~al.}(2006)\citenamefont {Wang},
  \citenamefont {Yates}, \citenamefont {Souza},\ and\ \citenamefont
  {Vanderbilt}}]{Wang_2006}%
  \BibitemOpen
  \bibfield  {author} {\bibinfo {author} {\bibfnamefont {X.}~\bibnamefont
  {Wang}}, \bibinfo {author} {\bibfnamefont {J.~R.}\ \bibnamefont {Yates}},
  \bibinfo {author} {\bibfnamefont {I.}~\bibnamefont {Souza}},\ and\ \bibinfo
  {author} {\bibfnamefont {D.}~\bibnamefont {Vanderbilt}},\ }\bibfield  {title}
  {\bibinfo {title} {{Ab initio calculation of the anomalous Hall conductivity
  by Wannier interpolation}},\ }\href
  {https://doi.org/10.1103/physrevb.74.195118} {\bibfield  {journal} {\bibinfo
  {journal} {Phys. Rev. B}\ }\textbf {\bibinfo {volume} {74}},\ \bibinfo
  {pages} {195118} (\bibinfo {year} {2006})}\BibitemShut {NoStop}%
\bibitem [{\citenamefont {Souza}\ \emph {et~al.}(2001)\citenamefont {Souza},
  \citenamefont {Marzari},\ and\ \citenamefont
  {Vanderbilt}}]{PhysRevB.65.035109}%
  \BibitemOpen
  \bibfield  {author} {\bibinfo {author} {\bibfnamefont {I.}~\bibnamefont
  {Souza}}, \bibinfo {author} {\bibfnamefont {N.}~\bibnamefont {Marzari}},\
  and\ \bibinfo {author} {\bibfnamefont {D.}~\bibnamefont {Vanderbilt}},\
  }\bibfield  {title} {\bibinfo {title} {{Maximally localized Wannier functions
  for entangled energy bands}},\ }\href
  {https://doi.org/10.1103/PhysRevB.65.035109} {\bibfield  {journal} {\bibinfo
  {journal} {Phys. Rev. B}\ }\textbf {\bibinfo {volume} {65}},\ \bibinfo
  {pages} {035109} (\bibinfo {year} {2001})}\BibitemShut {NoStop}%
\bibitem [{\citenamefont {Tsirkin}(2021)}]{Tsirkin:2021aa}%
  \BibitemOpen
  \bibfield  {author} {\bibinfo {author} {\bibfnamefont {S.~S.}\ \bibnamefont
  {Tsirkin}},\ }\bibfield  {title} {\bibinfo {title} {{High performance Wannier
  interpolation of Berry curvature and related quantities with WannierBerri
  code}},\ }\href {https://doi.org/10.1038/s41524-021-00498-5} {\bibfield
  {journal} {\bibinfo  {journal} {npj Comput. Mater.}\ }\textbf {\bibinfo
  {volume} {7}},\ \bibinfo {pages} {33} (\bibinfo {year} {2021})}\BibitemShut
  {NoStop}%
\bibitem [{Note1()}]{Note1}%
  \BibitemOpen
  \bibinfo {note} {See Figs. S1 and S2 of Supplemental Material at
  http://link.aps.org/supplemental/xxxxx . Spin resolved band structure with
  and without strains are shown. At $\Gamma $, the half-metallicity is robust
  while the size of band gap changes.}\BibitemShut {Stop}%
\bibitem [{Note2()}]{Note2}%
  \BibitemOpen
  \bibinfo {note} {See Fig. S8 of Supplemental Material at http://link.aps.org/supplemental/xxxxx. For this chemical potential shift, negative charge of 1.52 is needed.}\BibitemShut {Stop}%
 \bibitem [{Note3()}]{Note3}%
  \BibitemOpen
  \bibinfo {note} {Similar feature is found for opposite chemical potential shift, $\mu=E_F+0.12$ eV, where the magnitude of AHC monotonically decreases, or $\sigma_{xy}$ increases from -680 to -292 S/cm with strain. This is not by topological singularity but by band splitting associated with representation near $L$. See Sec. V of Supplemental Material at
  http://link.aps.org/supplemental/xxxxx.}\BibitemShut {Stop}%
\bibitem [{Note4()}]{Note4}%
  \BibitemOpen
  \bibinfo {note} {See Sec. III of Supplemental Material at
  http://link.aps.org/supplemental/xxxxx. Fermi surface sheets under different
  strains are shown for $+1\%$ interval, where the Fermi surface contour
  exhibits smoother and more continuous evolution.}\BibitemShut {Stop}%
\bibitem [{Note5()}]{Note5}%
  \BibitemOpen
  \bibinfo {note} {See Fig. S7 of Supplemental Material at
  http://link.aps.org/supplemental/xxxxx. Bands around $E_F$-0.3 eV, near nodal
  lines do not change with respect to strain.}\BibitemShut
  {Stop}%
\bibitem [{Note6()}]{Note6}%
  \BibitemOpen
  \bibinfo {note} {See Fig. S5, and S6 of Supplemental Material at
  http://link.aps.org/supplemental/xxxxx. For other strains, 
  the overall Berry curvature distribution is similar with different magnitudes.}\BibitemShut {Stop}%
\end{thebibliography}
%Control: production of eprint (1) enabled
%

\end{document}